%% file: ms.tex
\documentclass[12pt, preprint]{aastex}
\usepackage{emulateapj5}  

\def\asca{{\it ASCA }}		
\def\einstein{{\it Einstein }}	
\def\ergcm2s{~erg cm$^{-2}$ s$^{-1}$} 
\def\funit{~erg cm$^{-2}$ s$^{-1}$} 
\def\ergs{~erg s$^{-1}$}		
\def\cmsq{~cm$^{-2}$}		%
\def\nh{~$\rm{N_{H}}$}

\def\etal{et al.~}		
\def\kmsmpc{~km s$^{-1}$ Mpc$^{-1}$}		

\def\n4038{~NGC4038/39}		
\def\chandra{{\it Chandra }}
\def\x2{$\chi^{2}$}	



\begin{document}

\title{{\textit{CHANDRA}} Observations of ``The Antennae'' Galaxies (\n4038):
II. Detection and Analysis of Galaxian X-ray sources}

\author{ A. Zezas, G. Fabbiano, A. H. Rots, S. S. Murray}
\affil{Harvard-Smithsonian Center for Astrophysics,\\ 60 Garden
Street, Cambridge, MA 02138}
\shorttitle{\chandra\ Observations of ``The Antennae'' Galaxies II}
\shortauthors{Zezas et al.}
\bigskip

\begin{abstract}
We  report the detailed analysis of  the X-ray properties of the discrete sources
detected in a long (72~ks) \chandra ACIS-S observation of the Antennae
galaxies. We detect 49 sources, down to a detection limit of
$\sim10^{38}$~\ergs; 18 sources have $\rm{L_{X}>10^{39}}$~\ergs~
(Ultra Luminous X-ray sources; ULXs). Six of the 49 sources have an extended
component. Two sources show evidence for variability during this
observation, and three sources exhibit long term variability in
timescales of a few years. The spectra
of the discrete sources are  diverse, suggesting different emission
mechanisms, broadly correlated with the source luminosity: 
the most luminous sources exhibit harder emission, while 
the spectra of fainter sources appear softer.
Spectra and variability suggest that the ULXs may be binary
accretion sources; Supernova remnants or hot ISM in the \chandra
beam may be responsible for some of the softer sources.

\end{abstract}

\keywords{galaxies: peculiar --- galaxies: individual --- galaxies:
interactions --- X-rays: galaxies }

\section{Introduction}

The first imaging X-ray observations of nearby galaxies showed that a
large fraction of their X-ray emission is due to a population of
discrete X-ray sources:  Supernova remnants (SNRs) and X-ray binaries
(XRBs)  (e.g. Fabbiano 1989).
However, the properties of the X-ray source population and its dependence on 
galaxian parameters (such as star-formation) could not be investigated
before the deployment of \chandra because of the poor sensitivity and spatial resolution of the
 pre-\chandra X-ray observatories (e.g. Fabbiano 1995).

At a distance of  29~Mpc ($\rm{H_{O}=50}$\kmsmpc) the Antennae
(NGC~4038/39), the prototypical merging galaxies (Toomre \& Toomre 
1972), are a prime target for the study of
the X-ray source population in star-forming galaxies. 
The Antennae have been observed in the X-rays with almost every imaging 
X-ray telescope. The first \einstein observations detected a few
discrete sources as well as diffuse, possibly soft, emission
associated with this system  (Fabbiano \etal 1982, Fabbiano \&
Trinchieri 1983). Subsequent
observations with ROSAT HRI (with a resolution of $\sim5''$) resolved 12 discrete sources 
or emission regions
and yielded for the first time indications that some of them might be
Ultra Luminous X-ray sources (ULXs), emitting well in excess of the
Eddington limit for an accreting neutron star (Fabbiano \etal 1997). One of these sources was found
to be variable between two observations obtained one year
apart. Observations with the ROSAT PSPC and ASCA  (Read \etal
1995, Sansom \etal 1996) did not provide
significant information on the population of the discrete
sources, due to their poor spatial resolution. However, the ASCA
observations showed that there  is a 
significant hard X-ray component  (which could derive from a population of X-ray binaries 
 and/or from hot thermal gas), as well as a softer gaseous emission component.
 
NGC4038/39 were observed with \chandra ACIS-S, and the overall results
of the observations were reported by Fabbiano, Zezas \& Murray
(2001) (Paper~I). As reported there, a large population of extremely luminous
($\rm{L_{X} \sim 10^{38}-10^{40}}$~\ergs)
point-like sources is detected, along with a  diffuse component. 
The two components contribute equally to the total X-ray luminosity of
the system. Here, we  describe the detailed
analysis of the X-ray properties of the discrete 
sources.
These results will be elaborated upon and discussed in the
companion paper (Paper~III; Zezas, Fabbiano, Rots \&
Murray 2002), where  we also present a comparison with multiwavelength
properties of the emission regions.
 The derivation of the X-ray luminosity function (XLF) of the 
X-ray sources of the Antennae Galaxies
and a comparison with the  XLFs of other
galaxies is presented in Paper IV (Zezas \& Fabbiano 2002).

Throughout this paper we use a distance of 29~Mpc ($\rm{H_{0}=50}$\kmsmpc).
For $\rm{H_{0}=75}$\kmsmpc, the distance becomes 19.3~Mpc  
and all the cited luminosities will be a factor of 2.25 lower. At a
distance of 29~Mpc, 1~arcsec corresponds to a physical distance of
$140$~pc (at 19~Mpc it corresponds to 92~pc).
The quoted errors in the spectral
fits are at the 90\% confidence level for one interesting parameter,
unless otherwise stated. 

\section{Observations and Data Analysis}

This study is based on  observations (OBSID 315) obtained with the
ACIS-S camera (Garmire \etal 1997) on board the \chandra X-ray Observatory (Weisskopf \etal 2000). The
Antennae were observed for a total of 72~ks on December 1,
1999. 
 The first results from these observations have been reported in Paper~I.
The initial processing of the raw data was performed by the \chandra X-ray Center (CXC), with the
 reprocessing pipeline software (version R4CU5UPD6.5), using the  updated calibration files as of
September 2000.   In the following analysis we mainly used the CIAO
package \footnote{http://cxc.harvard.edu/ciao}. The geral data
analysis techniques are presented in the \chandra Data Analysis
Threads \footnote{http://cxc.harvard.edu/ciao/documents\_threads.html}.
For our analysis we used the Level-2 event
files  which include only events of
\asca grades 0,2,3,4,6 and status 0.
We searched the data  for periods of enhanced
background  radiation by extracting a light curve from a large
source-free region of the ACIS-S3 chip. 
The background as measured from the event2 file (prior to any processing
and including events of all energies) was found to be fairly constant,
at a level of 
$0.456 \pm 0.004$ counts/sec in an area of 7.06 arcmin$^2$. This is
roughly half of the quiescence background level reported in the
\chandra Proposers' Observatory Guide.  We  screened the
raw data to exclude events with  energies outside the (0.3 - 10)~keV  band. The 0.3~keV cut-off is chosen
because of the on-board event rejection, whereas the 10.0 keV cutoff
is set because the telescope's High Resolution Mirror Assembly (HRMA;
Van Speybroeck \etal 1997) 
 effective area  is effectively zero above 10~keV.  The ACIS-S3 chip which
was in the focal point of the HRMA has two
``hot columns'' at  the borders of each node, which we excluded (see Paper I).   

In order to check the absolute astrometry of these observations, we
searched the U.S. Naval Observatory (USNO) star catalogue for stars
within 4 arcmin from
the galaxy. We found three stars with X-ray counterparts. The
offsets between the optical and the X-ray sources were not systematic and less than $0.5\arcsec$. 
Therefore we conclude that the absolute astrometry
of these observations is good to within $0.5''$, in accordance
with  estimates from the calibration team (Aldcroft \etal 2001)\footnote{http://cxc.harvard.edu/cal/ASPECT/celmon/}.

\subsection{ Source Detection}

After the initial processing we created images in the (0.3-2.0)~keV, (2.0-4.0)~keV,
and (4.0-10.0)~keV bands. The lower energy band  allows direct comparison 
with the ROSAT results. We decided to divide the (2.0-10.0)~keV band in two 
intervals at 4.0~keV, in order to  obtain more information on the spectral shape of the X-ray emission.
 The  4.0~keV boundary was chosen
because at higher energies the effective area of the ACIS-S3 chip begins to drop and we wanted to retain 
a relatively good signal to noise ratio (S/N) for the objects detected in the hard band.  We
smoothed each image using the adaptive smoothing algorithm implemented
in the {\textit{csmooth}} tool of the CIAO v2.0 suite. This algorithm
adjusts the size of the smoothing kernel (in our case a two-dimensional Gaussian)
in order to maintain a uniform S/N ratio over the smoothed image. For
these images the S/N ratio was set to be between 3 and 5. The
smoothing scales range between 0.5 arcsec 
(for strong point sources) and 256 arcsec (for very low surface
brightness diffuse emission). The majority of the  
image is smoothed with a Gaussian with a FWHM of $\sim15$~pixels. The images
in the three energy bands, together with a full band unsmoothed image, are presented in Fig. 1.   

  We searched the unsmoothed images in the three energy bands for
discrete sources, using both
 the sliding cell and wavelet algorithms implemented in the
CIAO {\textit{celldetect}} and {\textit{wavdetect}}  tools, respectively.
We ran {\textit{celldetect}} with a cell corresponding to 90\% 
of the encircled energy of a point source which varied with  off-axis angle. The
cell size ranges between 6 pixels for an on-axis source to 12 pixels
for a source at 2 arcmin off-axis (which is the maximum in the case
of the Antennae). The background was estimated from an annulus around the
source defined to cover the same area as the source cell. Because the detection
cell is not adjustable to the size of the source, the effectiveness of
this algorithm  is limited to unconfused regions 
and  point-like sources. It also strongly depends on the local
background, which may well include other nearby sources.  The
3$\sigma$ best fit ellipses to
the spatial distribution of the events for each source are plotted as
red ellipses  in Fig. 1
on the raw (0.3-10.0)~keV band image.

As the X-ray emission of the Antennae is very patchy, with crowded regions of
 sources embedded in diffuse emission, we also used the
{\textit{wavedect}} source detection tool which performs better in these
``difficult'' conditions (Freeman \etal 2001a). The advantage of this tool
 is that it parameterizes the shape of a point-like source with
the  ``Mexican Hat'' (MH) function,  consisting of a narrow core and
relatively extended wings. This function can be spatially rescaled to
match the extent of each source, allowing one to obtain information on the spatial properties
of the detected sources and to detect and accurately measure fluxes for 
sources at various spatial
scales.  In our analysis,  we searched for sources on scales of 1, 2,
4, 8 and 16  pixels. These scales correspond to the radius R of the
core of the MH function (its Full Width at Zero Intensity is
$2\sqrt{2}\times R)$. The determination of the background was done 
from a background map produced by counts detected in the wings of the
MH function (see the {\textit{Detect}} guide for more 
details; Dobrzycki \etal 2000).
  
 In order to establish whether a source is point-like or extended and to
estimate its total emission (corrected for the Point Spread Function;
PSF), {\textit{wavedetect}} compares the scale-length which matches best the shape of
the source with the size of the PSF at its position.
Below 4.0~keV we used models of the PSF at 1.49~keV as most of the
photons are detected between 1.0-2.0~keV,  where the ACIS-S effective
area peaks. Moreover, at this energy 
the HRMA PSF is most accurately calibrated.
For higher energies  we used the calibration at 4.5~keV since above this
energy  the effective area of the detector falls rapidly and even for
the hardest sources most of the photons are detected below
6.0~keV. The limiting  chance detection probability was set to
$10^{-6}$, which corresponds to $\sim0.5$ spurious sources in the
searched area (Dobrzycki \etal 2000).  Fig. 1a shows the 3$\sigma$  ellipse fit to each
source found by {\textit{wavdetect}} as white ellipses.

The merged source list including results from both {\textit{celldetect}} and {\textit{wavdetect}}
results in the (0.3-10.0)~keV band is presented in Table 1. For
completeness we also include in this table   sources detected only in the
soft or medium band.  Column (1)
gives a number indentifying the sources internally in this paper (sources 1-49), sources 1c, 2c and 3c
 at the bottom of the table are sources detected only by {\textit{ celldetect}}, 
Column (2) gives the 
CXO source name, Columns (3) and (4) give the RA and Dec (J2000),
Column (5) gives the background subtracted source counts together with
the $1\sigma$ statistical error following the Gehrels approximation
for low count statistics
(Gehrels 1986), Columns (6) and (7) list the background contribution to
the total source counts in the detection region of each source and the S/N
 ratio  of the detection, Column (8) gives
the extent of each source in terms of the fraction of the PSF size at
its position on the detector (defined as the
geometrical mean of the  minor and major axis of the $1\sigma$ ellipse
divided by the 39\% encircled energy radius of a point source in the soft (0.3-2.0~keV band)).  
 Columns (9) to (14)
give similar information as Columns (3)-(9) but for the sources  detected
with {\textit{celldetect}}. Columns (15) and (16)  give the logarithm of the
observed and absorption-corrected (0.1-10.0)~keV band luminosity
respectively  for each source
in ~\ergs~ (assuming a 5~keV bremsstrahlung model
 and galactic line of sight absorption
$\rm{N_{H}=3.4\times10^{20}cm^{-2}}$; Stark \etal 1992, and using the
AO2 release of the ACIS-S3 effective area). 
 For sources detected with both
algorithms these luminosities are based on the count rate measured
with {\textit{wavdetect}}, since it estimates more accurately the
source intensity and the local background. For sources detected only in the soft band
we estimate the total band luminosity assuming the same spectrum as
for sources detected in the full band. However, for sources detected
only the medium band we also use a 5~keV thermal bremsstrahlung model
but with a column density of $2\times10^{21}$\cmsq~ based on spectral
fitting results (section 2.4). 
Table~1 shows that for the sources
detected with both {\textit{celldetect}} and {\textit{wavdetect}} the derived 
parameters agree fairly well:  
the measured net source counts  are typically consistent within 30\%.
The greatest differences are for resolved sources which are slighlty
extended or confused and therefore not
well  modeled with the detection cell used by {\textit{celldetect}}.

Fig. 1a shows the position of each source detected with {\textit{wavdetect}} or
{\textit{celldetect}} (marked by the 3$\sigma$ ellipse) 
 overlaid on the full-band raw data. The sources
are numbered  following the
 convention of Table 1. As it was expected, {\textit{wavdetect}}
is more efficient than {\textit{celldetect}}, in detecting
and separating close sources.   This
figure also shows that the three sources detected with
{\textit{celldetect}}  alone and listed at the bottom of Table~1 are local
enhancements of the diffuse 
emission which are rejected by the source
selection algorithm of {\textit{wavdetect}}, and therefore we do not
include them in our following analysis.

 In Table 2 we present the results from the {\textit{wavdetect}} run
 for each energy band separately. Column (1) has the source identification  number used in this
paper, columns (2) to (5) give the net source counts (with 1$\sigma$ statistical
error), the background counts in the  source 
area, the S/N  of the source (in $\sigma$) and its extent (defined in
 as in Table~1). Columns (6)-(9) and (10)-(13) give the same
information for the medium (2.0-4.0 ~keV) and hard (4.0-10.0 keV) bands.  
 In both Tables 1 and 2 we included only sources detected at the 3$\sigma$ level above
the local background and   contained within the optical outline of the
Antennae galaxies. Figs. 1b,c,d show the position of each source
(marked by their $3\sigma$ best fit ellipse in the full band), following the
numbering convention of Table 1, overlaid on adaptively  smoothed soft, medium and hard
band images.

 We detect a total of 49 sources, possibly associated with the Antennae.
The limiting luminosity is  $\sim10^{38}$\ergs~ ($\rm{H_{0}=50}$\kmsmpc;
$\sim5\times10^{37}$\ergs, for  $\rm{H_{0}=75}$\kmsmpc). 31 sources (38 for
$\rm{H_{0}=75}$\kmsmpc) have luminosities below $10^{39}$\ergs~ 
while 18 (11) have higher luminosities, so, as reported in Paper~I, they are Ultraluminous
X-ray Sources (ULXs), emitting well in excess of the Eddington limit
of a neutron star accretion binary ($\sim 3 \times 10^{38} \rm ergs~s^{-1}$). 
The  slight difference
between these results and those presented in Paper~I are due to the
use of more appropriate effective areas and unabsorbed fluxes for the
determination of the luminosities.
 To estimate the  number of  sources not  physically associated with
the Antennae we performed  a
 simple calculation using both the logN$-$logS relation of Giacconi \etal
(2000) and source counts of background sources from this observation.
We found  that only 2.6 serendipitous sources are expected in the area of
the Antennae, down to a luminosity level of $1\times10^{38}$~\ergs~ (Paper I).
This number reduces to 0.6 sources for sources
brighter than $10^{39}$~\ergs. We therefore conclude that most of the 
detected sources, and especially the higher luminosity ones, do belong to NGC4038/39.

\subsection{Spatial Analysis}

Determining if a source is point-like or extended depends on the S/N
of the source, because this limits the accuracy of our measurements
of the spatial distribution of source counts.
We used 100 counts as a threshold for our investigation of source
extent. A more detailed analysis of the extent of fainter sources will
be presented in Paper V (Fabbiano \etal 2002, in prep). Due to the
very high spatial resolution of \chandra,sources with fewer counts
are detected over typically a $2\times2$ or $3\times3$ pixel cell, which is
too small to allow any spatial profile fitting. Spatial fitting is also
difficult because the ACIS undersamples the HRMA PSF 
(the 50\% encircled energy radius of HRMA at the focal aim-point is
$\sim0.3''$ while the ACIS pixel size is $0.49''$).
For this reason we consider all the sources with fewer than 100 counts
and small source ellipses in the (0.1-10.0)~keV band as point-like,
without doing any profile fitting. Instead we obtained
radial profiles for all sources detected with more than 100 counts. We
 also derived radial profile for four sources
 detected with fewer than 100 counts, but having $3\sigma$ best
fit ellipses with major axis larger than 8~pixels (sources 5,6,7 and
10), in order to investigate if they have a point-like  conponent.

For these sources we created
models of the PSF, appropriate to the position of each source on the detector, using
the CIAO {\textit{mkpsf}} tool which generates a PSF 
 at a given energy and off-axis angle, based on
pre-flight calibration. As the best calibration
of the PSF of ACIS is at 
the energy of 1.459~keV, where also the sensitivity of ACIS-S3 is
optimized, we extracted our model PSFs at this energy and we compared
them with the respective  source  
profiles in the 0.3-2.0~keV band. Although these model PSFs do not
take  into account the telescope motion during the observation, the residuals
after the application of the aspect solution are usually less than
$\sim0.75''$ ($1\sigma$ level)  
\footnote{http://asc.harvard.edu/mta/ASPECT/celmon/index.html}, 
and therefore we can directly compare these models with
the actual observations.
 After calculating the model PSFs, we rescaled them to the total number of photons
detected in each source. The PSF becomes  asymmetric at large off-axis angles
($\theta\geq1.5'$). This mainly affects the sources associated with
NGC~4039  (Southern galaxy) since the optical axis was roughly on the
position of the nucleus of NGC4038 (Northern galaxy).
   For this reason we used elliptical annuli instead of
circular annuli for the profile extraction.  
 The PSF consists  of a narrow core and  more extended wings, 
which rotate depending 
on the  position of the source relative to the aimpoint, thus making the
determination of the position angle of the  
major axis non-trivial. For our analysis we determined the angle of
the major axis directly from the model PSF.

 We used the CIAO application $\textit{Sherpa}$ (Freeman \etal 2001b)
to fit the radial profiles of the sources
using as models the radial profiles of the simulated PSFs (extracted
using exactly the same regions as for the source profiles). In the
cases which show residual wings, due to incomplete subtraction of the source
background we added  a constant to the PSF.  For each source,
identified in Column (1),  Table 3 presents the
results of these fits:  Column
(2) gives the \x2 together with the number of degrees of freedom for
each fit with the model PSF,  and Column (3) gives the same
information for fits with the PSF plus a constant background. 
Figure 2 shows the results of the spectral fits with the PSF plus
background model for 3 examples of point-like sources and for all the
extended sources.
The points display the source profile, whereas the solid line represents
the model PSF. The bottom panel of each plot shows the fit residuals
in $\sigma$. The fits with the PSF alone are in most cases very poor
and the inclusion of a constant improves them in the 99\% confidence level.

In order to obtain a measurement of the physical scale of each source we also
fitted both the source and the model PSFs with a Gaussian. The results of
these fits are also presented in Table 3, where Columns (4) and (6)
give the FWHM of the source and the model PSF profile in ACIS-S pixels (with $1\sigma$
errors for one interesting parameter), and Columns (5)
and (7) give the \x2 and number of d.o.f for the source and the
model PSF respectively. For some sources the resulting reduced \x2 is
vey high. This is due to residual local diffuse emission  and also to the wings of the PSF which are not well
modeled by a single Gaussian. For this reason we fitted the profiles
with a Gaussian and a constant.  These results are presented in 
 Columns (8)-(11). In Column (12) we give the deconvolved FWHM of the
 Gaussian profile of the extended sources
in arcseconds and parsecs (assuming a distance of 29~Mpc), point like
sources are marked as ``p''. In general the latter models gave a
statistically significant improvement in the fit  based on an F-test
with 1 additional parameter.  
We find that all but 6 sources for which we measured the profiles  
are consistent with the model PSFs and
therefore we consider them as point-like. The FWHM of the Gaussian
used to fit the profiles of sources 5, 6, 7, 10, 24 and 29 are larger than
those of the PSF, above the $3\sigma$ confidence level (source 10 has
a weak point-like core but very extended wings and we consider it as
extended).  This clearly
suggests that these sources have an extended component. Their extent ranges between
$3''$ and $15''$ corresponding to a physical scale of $\sim400$~pc to
$\sim2.1$~kpc, as measured by the FWHM of the best fit Gaussian. 
Sources 25 and 29 correspond to the Northern and 
Southern nuclear regions respectively, as determined by their radio
positions (Neff \& Ulvestand, 2001).

\subsection{Timing Analysis}

Because of the low count rates of the sources in the Antennae, it is very hard to
search for short term variability.  We extracted
light-curves for the 5 brightest sources (net count rate $> 0.001~\rm{counts/sec})$ 
binned with bin sizes of  500~s, 1000~s and 5000~s and we 
compared them with the standard deviation of the mean count rate for
each light-curve. In all cases the data points were consistent with
the respective mean values at the
99\% confidence level.
We also split the observation in five exposures of 15~ks. Again no
variability was detected.

We also searched for variability in all the sources by comparing
the cumulative distribution of the photon 
arrival times for each one  with the photon arrival times of events in
a large source free background region around the galaxy. In order to
minimize the background we only used events in the (0.3-7.0)~keV band,
because the particle background dominates above 7~keV.
 The comparison was performed using the
Kolmogorov-Smirnov test (KS test). This test showed that for the vast
majority of the sources
the two distributions are consistent with being drawn form the same parent population,
i.e., the sources are not variable at the 99\% significance
level.  Only for sources 14 and 44 we detect variability at a
significance level of 99.2\% and 99.7\%, respectively.  Source 14 is a
relatively bright source in the southern part of the galaxy. A
 light-curve of this source in bins of 15ks suggests a decline of its flux,
although because of the  large error bars  a constant
still gives an acceptable fit. Source
44 is the third brightest source in the Antennae. A light curve in
bins of 15ks suggests that there is a flare in the second bin, but again
because of the low S/N a constant gives a good fit. Fig. 3 shows the
cumulative distributions of the two sources (solid lines) compared
with the distributions of the background (dashed lines). A similar
test in the (2.0-10.0)~keV band showed no variability signs for any 
sources. This could be caused by  the low number of counts for most
sources above 2~keV.

 In order to search for long term variability we compared our data
with the ROSAT HRI observations of the Antennae (Fabbiano \etal
1997). Prior to \chandra these are the highest
angular resolution observations in existence. In
order to directly compare our data with the HRI measurements, we concentrated on
isolated sources (X-3, X-4, X-8, X-11, X-12, X-13 from Fabbiano \etal
1997). We
also used exactly the same regions used to measure the HRI
luminosities;  we measured  background from
regions  outside the galaxy. In order to match the HRI band we used the (0.3-2.0)~keV \chandra images
to measure the count rates of the sources. Then we converted both the
HRI and the Chandra count rates to luminosities in the (0.1-2.5)~keV band assuming a 5~keV
thermal bremsstrahlung model with Galactic line-of-sight absorption
($3.4\times10^{20}$\cmsq). These results are given in Table 4, where
Column (1) gives the HRI source, Column (2) the  \chandra
source(s) encompassed by the extraction region, Column (3) the HRI
luminosity,  Column (4) the \chandra luminosity, and Column (5) the
significance (in $\sigma$) of the difference between the ROSAT and
\chandra measurements.
These results suggest that sources 16, 42 and the combination of sources 44 and 46 show
variability or a factor of $\sim2$ in timescales of a few years.


\subsection{Spectral Analysis}

As mentioned in the introduction, there are two major components in the
galaxian source populations: X-ray binaries and SNRs. X-ray spectra
are a very powerfull tool for distinguishing between these two types of
sources. Moreover, from  the parameters of detailed spectral fitting it
is possible to distinguish between different types of X-ray binaries
(e.g. binaries in different states, pulsar from black-hole binaries,
see van Paradijs 1999). 

We extracted spectra for all the detected sources, using extraction regions
 defined  to include as many of the source photons as possible, but at the same
time minimizing contamination  from nearby sources and background. The 
background region was usually a source-free circular or elliptical
annulus surrounding
 each source,  in order to take into account the spatial
variations of the  diffuse emission and to minimize effects related to
the spatial variations of the CCD response. Any sources encompassed by
the background region were excluded. The regions used to extract the
source and background spectra are presented in Fig. 4a and 4b.
 In order to take into  account spatial variations of the detector
gain, the spectra were extracted in PI (Pulse Invariant; gain-corrected 
space.

For each spectrum we created response matrices and ancillary
response matrices using the {\textit{mkrmf}} and {\textit{mkarf}} tools in the CIAO v.2.0
package with the N0001 (December 2000) release of the response
calibration files (in the form of FITS Embedded Functions; FEFs).
 The variations of the charge transfer inefficiency (CTI) in adjacent  
FEF regions of the ACIS-S3
chip, within which the
response is modeled and assumed to remain stable,
are very small. For the case of ACIS-S3 chip which was at the focus of the telescope,
these regions are 32$\times$32 pixels large.  
For point sources which encompass
more than one FEF regions, we extracted response matrices from the FEF
which includes most of the source photons. In order to assess
 the uncertainty of this approximation, we also 
created an RMF for the source which encompasses the largest number of
FEF regions, using the {\textit{calcrmf}} tool \footnote{http://asc.harvard.edu\   }. 
 This tool calculates the response matrix for a  source by
combining the appropriate FEFs corresponding to the detector pixels on
which source photons are detected after taking into account the aspect
solution. The FEFs are then  weighted 
 by the number of counts in each region. We find that the difference
in the RMFs calculated by the two different  
methods are below the 10\% 
level, giving us confidence that the use of a single FEF in
the calculation of the RMF does not affect our results.  In the case
of a few sources which  fall at the boundary between two ACIS nodes we used the response
matrix  for the node onto which most of the events are detected. Since
these sources have typically less that 200 counts, this will not affect
our spectral fitting results, which are dominated by source statistics.

 The spectra were fitted using the XSPEC~v10.0 package (Arnaud \etal 1996). 
In order to use \x2 statistics, we grouped the data to include at least  
15 counts per spectral bin before background subtraction.  In the spectral
fitting, we excluded  any events with energies above
10.0~keV or below 0.3~keV. 
Although, the calibration at energies below 1.0~keV is
uncertain, we retained data in the
this range because the statistical error of the data 
points at
these energies is still significantly larger than the expected calibration
uncertainty.
Moreover, these data are important for 
constraining the absorbing column density. 
It is known that   the ''energy scale'' has a shift of 
$\sim20$~eV \footnote{http://asc.harvard.edu/cal/Links/Acis/acis/Cal\_projects/Energyscale\_120.html}. Although this feature is not taken into account
in the release of the calibration data we used, it does not affect
our analysis as each spectral bin below 1.0~keV covers an energy
range much larger than the energy resolution of ACIS-S3. 

\subsubsection{Single component models}

We fitted the spectra of all the sources with more than $\sim50$~net counts (5 spectral bins or more)
 in the 0.3-10.0keV band,  with simple one
component models: absorbed power-law (PO) and absorbed Raymond-Smith
(RS; Raymond \& Smith 1977). In both cases, the $\rm{N_{H}}$  was initialy fixed to
the Galactic ($3.4\times10^{20}$\cmsq) value. 
 The RS model was suggested by the residuals around 1~keV in the PO
fits of some sources, as a thermal plasma component produces the Fe-K$\alpha$ blend in
this energy region.  
We then left the $\rm{N_{H}}$ free to vary and found an improved  fit
in all cases at the 99\% confidence level, based on an F-test for one
additional parameter. The best fit $\rm{N_{H}}$  is typically  
higher than the Galactic value. In all the fits with the RS model the
abundance was fixed to the solar value. 
The results of the fits are summarized in Table 5: Column (1)
gives the source identifier,  Columns (2) and (3) the photon index and the \x2
(with the number of degrees of freedom; dof) for the PO model with
absorption fixed to the galactic. Columns (4)-(6) give the results for
the fits with the PO model and free \nh: Col. (4) gives the  photon index
$\Gamma$,  Col. (5) gives the  best fit 
absorbing column density in units of $10^{22}$\cmsq, and Col (6) gives the \x2 with the dof. Columns
(7) and (11) give the analogous results for the RS models. All errors
are at the 90\% confidence level for one parameter of interest.  The
spectra, together with the best fit PO models and the resulting
residuals are presented in Figure
5. In the same figure we present 
$\rm{\Gamma-N_{H}}$ confidence contours for these fits. The contours are at the 1,2 and
3$\sigma$ levels for two  parameters of interest. 

In most cases the single component PO and RS  models give satisfactory
fits.  Only for 10 sources it is possible to favor one of the two models,
 based on an F-test. The \x2 of these sources 
 is shown in bold type in Table~5.   We note that for a few sources
the PO fits give relatively large best fit absorbing column
densities. For some of them this is clearly indicated by the low
energy cutoff of their spectra (e.g. sources 12, 24,25, 33, 35;
Fig.~5).  However,  for other sources 
  the high inferred absorption may be due to inadequate fitting of the
continuum by the PO model as for example in the 3 sources (src 6, 10,
18) which are
better fitted with an RS model, where the derived column density is
significantly lower than that obtained by the PO fits.  
In the next section we will discuss more complex models which  give in most cases
 lower overall absorption but usually 
require a heavily absorbed hard component.

In Table 6 we give the luminosities of the individual sources based
on the best fit PO model with free \nh~ except for the sources which can be
modeled  better or equally well 
 with a complex model (\S2.4.2) where we used the latter to determine the
luminosities. In this Table Column (1)
gives the source number, Column (2) gives the background subtracted
count rate in the (0.3-10.0)~keV band, Columns (3) and (4) give the
observed flux in the soft (0.1-2.5)~keV and hard (2.5-10.0)~keV bands
whereas Columns (5) and (6) give the flux corrected for the line of
sight Galactic absorption (but keeping the remaining intrinsic
absorption) and Columns (7), (8) give the flux corrected for the total
absorption in the same bands. The fluxes are given in units of
$10^{-14}$\funit. Columns (9)-(14) give the logarithm of the observed
and absorption corrected  luminosities in the soft and hard bands (in
units of \ergs). Although these are the most accurate estimates of the
source luminosities, they are not available for all the sources; for
this reason throughout the paper we will use luminosities from Table~1
unless otherwise stated.

\subsubsection{Complex models}

 From the data residuals relative to the best fit  power-law models
(Fig. 5), one can see that there is a large number of sources with a
soft excess below 2.0~keV.  For this  reason, the spectra of the
sources with adequate numbers of counts were fitted with a double
component model consisting of a PO+RS spectrum. The PO is seen through
an absorber free to vary, whereas the total spectrum is observed
through the Galactic line of sight \nh~(fixed).  
The results of these fits are  presented in Table 7.
In this Table Column (1) give the source number, column (2) and
(3) give the best fit temperature and photon index for the RS and the
PO models respectively and column (4) gives the  column density in
units of $10^{22}\rm{cm^{-2}}$. Column (5)  gives the \x2
and the number of dof.   In three cases only, is the
improvement in the fit over the single component models 
statistically significant above the 99\% confidence level, based on an
F-test for two additional parameters (the abundance of the RS was
fixed to the solar value as was for the single component models). The
\x2  for these sources are shown in bold. For these sources (src
5,6,29),  the resulting photon index is much steeper than with the
PO models alone. This is expected since now the soft excess if modeled by
the additional thermal component. The temperatures of this thermal
component are generally between 0.3 and 1.0~keV, 
with a few sources having temperatures up to 3.0~keV.  
Although with the current data we cannot discriminate between one and
two component models for most sources, the PO+RS  fits indicate that in 5
cases there may be a highly obscured (\nh$>0.3$\cmsq, 10 times higher
than the Galactic column) hard component. 

In order to constrain any  excess  absorption for the
hard component and assess its significance we fitted the spectra using the following method.   
First we fitted the spectra above 2.0~keV or 1.5~keV (depending on the
number of bins in the spectrum) with a single unabsorbed power-law. Then we fitted the
 full band spectrum with a PO and RS model with the photon index fixed to 
the previously  obtained  value and the other parameters  free to
vary. The results from this fit  for the sources for which it was possible (i.e.
those  with more than 3 bins above 1.5keV) are presented in Table~7:  
in Column (6) we give the photon index (determined from the fit
above 2.0~keV), in Columns (7) and  (8)  we give the temperature of
the RS component and the overall absorption (in units of
$10^{22}\rm{cm^{-2}}$)  and in Column (9) we give
the \x2 and the number of dof for this fit.  Then for the sources
which gave in the previous fit a reduced \x2 higher than 1.0,  we included  a
second absorber applying it only to the PO component
and we fitted the spectrum again. For this fit  the photon
index  of the PO component was also fixed to the previous
values. 
 These results  are also presented in Table~7.
 In Column (10) we give the the best fit temperature for the thermal
component and in Columns (11) and (12) the intrinsic absorption of the
hard component and  the overal absorbing column density. Finally in Column (13) 
 we give the \x2 and the number of d.o.f for the fits 
after adding the second absorber. Although generally the
 best fit absorption of the PO component is higher than the
overall absorption,  for none of the sources  the improvement of the fit is statistically
significant. 
However, the PO model for some sources predicts unrealistically high
luminosities ($\rm{L_{X}>10^{42}}$~\ergs), while  the double component
models predict more realistic luminosities. This suggests that the latter
may be a more realistic model, although	 keeping in mind that given the
limitations of the present data we have no proof of it. 

Fits with more complex models such as
power-law plus a single or multi-temperature black-body component,
appropriate to model the  emission 
of X-ray binaries (e.g. Nagase 1989), did not yield
 a statistically significant improvement in the fits (above the
99\% confidence level based on an F-test).

\subsubsection{Line features}

The residuals after the PO fits show evidence for
emission lines in the spectra of sources 11, 16, 37, 42 and 45
(Fig. 5). We therefore 
modeled the spectrum of these sources with an 
absorbed power-law model to which we added narrow gaussians to account
for the emission lines. 
 The addition of these Gaussian components is
significant at the 99\% confidence level in three cases (sources 11, 16
and 42). The parameters
of the lines for these three sources are presented in Table 8. In this Table Column (1)
gives the source ID, Column (2) gives the energy of the line, Column
(3) gives the identification of the line, Column (4) gives the
normalization of the Gaussian and Column (5) gives the  \x2 and the
number of degrees of freedom. Regarding the identification of the
lines, we note that these strongly depend on the exact energy of the
line. Therefore, although the emitting element is well identified, there
is some uncertainty in the identification of the exact species which
is reflected in the ranges given in column (3) of Table 8.
Surprisingly, in every spectrum we find emission lines from different
species suggesting that the gas parameters and the environment of the
sources are very diverse. Inspection of high S/N ACIS-S3 background spectra
(Markevitch \etal
2001)\footnote{http://hea-www.harvard.edu/~maxim/axaf/bg/index.html} 
show that there are two very strong emission
lines at $\sim1.7$~keV and $\sim2.2$~keV.
Only the  line we detect in source~16 at 2.17~keV could be due to residual background.
All other lines are likely to originate from the source. In Fig. 6 we show the 
spectra of these three sources together with a PO model of the
continuum radiation and the
 Gaussians used to model the emission lines. The bottom panel of the spectra 
 shows the ratio of the data and the model.

\subsubsection{Hardness ratios}

There are 19 sources with less than 50 net counts,  for which it is not possible to 
perform any spectral fitting, even with the simplest models. For this reason we 
tried  to determine their spectral parameters by using hardness ratios (HR). We 
calculated  hardness ratios using the (0.3-1.0)~keV (soft; S), 
(1.0-2.5)~keV (medium; M) and (2.5-7.0)~keV (hard; H) bands. We selected
these bands after  detailed modeling  
which showed that these are optimal  to distinguish between
different spectral models, when combined in HR-HR diagrams. 
We define  three hardness ratios as: $\rm{HR1=(S-H)/(S+H)}$,
$\rm{HR2=(S-M)/(S+M)}$ and $\rm{HR3=(M-H)/(M+H)}$. 
The HRs were calculated by extracting the unbinned
spectra  and binning them  
in the S, M, and H  energy bands. To subtract the background we followed
the same procedure but for annular regions 
around the sources.  Then the background was rescaled to the area of
the source extraction cell and sutbracted from the source counts in
each band.

To calibrate  the HR-HR diagrams, we  calculated
hardness ratios from simulated  absorbed power-law and
thermal plasma spectra, for a wide range of  parameters. We used the 
response matrix for an on-axis point source. Based on these model
spectra we created grids on plots involving two different hardness
ratios.  We checked the accuracy of these grids by comparing them with
the  hardness ratios of  sources with good spectral fit results (see
Fig.~7).   In this figure the red grids correspond to the PO models and the green 
grids correspond to the RS models. In the sides of the grids we give the values of the
model parameters. The column density in both cases ranges between 0.01$\times10^{22}$\cmsq~ and 
0.5$\times10^{22}$\cmsq~
((0.01, 0.025,  0.05, 0.075, 0.10, 0.12, 0.25,
0.5)$\times10^{22}$\cmsq). The temperature ranges between 2.0 and 10.0~keV  
(for lower temperatures these grids become degenerate and is not possible
to distinguish between different temperature), and the photon index ranges between 0.5 and 2.0.
 In general the agreement between the spectral parameters 
derived from spectral fits and those estimated from
the HR-HR diagrams is very good. Only weak sources 
with a soft excess appear to have much lower  \nh~ than derived by spectral
fitting. This is because the determination of  \nh~from
hardness ratios depends on the S-M ratio. If a source has even a small
soft excess, this results in a larger S-M ratio and therefore a smaller
estimated \nh.  
The estimated spectral parameters are consistent between the 
different diagrams. The HR1-HR3 diagram is best suited for determining
the spectral parameters, because of the  
larger separation of the hardness ratios for different spectral
models. This is very important in the case of very faint sources where
the  error bars in the HRs are 
very large.  Fig.~8 shows the hardness ratio diagrams for the faint sources. We were able to 
apply these diagrams to 16 sources which  were detected  in at least two of the three bands. 
  This figure shows that 
the majority of the sources have soft spectra. Also some of them may have an additional 
soft thermal component.

\subsection{Obscured sources}

 Spectral fits with a single power-law suggested that 12 sources are
observed through a column density at least 10 times higher than that
of the Galactic  line of sight \nh~ (Table~5). However, for two of
them (src 10, 18) a thermal
model gives a better fit with a lower column density (constistent with
the Galactic \nh). For three other sources (src 5, 6, 29) a composite thermal/power-law
model also gives a better fit with an absorption much lower than that
required by the single power-law model.  For one additional source
(src~40) the composite model also gives a significantly lower \nh~
although the improvement in the fit was not statistically
significant. We consider this source as non obscured since its
spectrum (Fig~5) does not show any significant cutoff below
1.0~keV. For the other sources for which it was possible to perform
more complicated fits, the  composite models indicate
absorption for the hard component slightly lower than the overal absorption measured
with the single PO model (Table~7), although the improvent in the fit over the
latter is not statistically significant. 
  Therefore, we conclude that we have 6 highly obscured sources
(i.e. \nh$>0.34\times10^{22}$\cmsq): sources 12, 24, 25, 34, 35 and
36.

\subsection{Spectral trends}

The spectra of X-ray sources are  useful tools for understanding
 their nature, since different types
 of objects have different spectral signatures.
Pulsar XRBs as well as BH binaries in  low state tend to have hard spectra  
($\rm{\Gamma<2.0}$)
 (e.g. Nagase 1989, Tanaka \& Lewin 1995). The spectra of BH binaries in  high state
 and unmagnetised neutron stars are dominated by a multi-temperature
disk-BB spectrum (e.g. van Paradijs 1999,
 Tanaka \& Lewin 1995).  SNRs  typically exhibit soft thermal emission 
(kT~$\sim0.5-5.0$~keV)
 (Schlegel 1995), although young supernovae detonating in dense
environments (cSNRs)   can have much harder spectra (Plewa 1995, Terlevich 1994).

From both the spectral fit results and the HR diagrams, it is clear that
the discrete sources in the Antennae cover a very wide range of
spectral parameters. Of course this is not surprising for such a large
number of sources spanning 3 orders of magnitude in luminosity,
 and for an actively star-forming galaxy, which may
contain several generations of stellar objects.
We searched for possible trends in the spectral shape, in order to
 understand which types of objects dominate in each luminosity range. First we
plotted the source luminosity (corrected for galactic
absorption) against the best fit photon index   from the single
component model (Fig.~9). We plot the
point like sources in red,  and the extended
sources in blue. The typical uncertainties in the luminosity are
$\sim5\%$ for the most luminous sources and $\sim20\%$ for the
faintest.  We see a trend for more luminous sources to have harder
spectra. When we split the sample in three subsamples containing
$\sim10$ sources each,  we find that  the mean photon index for the
first bin ($\rm{L_{X}<5\times10^{38}}$~\ergs) is $6.89\pm2.15$, for
the second   ($\rm{5\times10^{38}<L_{X}<10^{39}}$~\ergs)
  $2.19\pm0.62$, and for the third ($\rm{L_{X}>10^{39}}$~\ergs)
  $1.56\pm0.27$. These are represented by the black points with the error
bars in the bottom of Fig.~9. These results suggest that
there may be a marginal statistically
significant trend for more luminous sources to have harder spectra
 ($\sim2\sigma$ difference between the faintest and
brightest sources). 

A  plot of the
X-ray luminosity against \nh~ (also in Fig.~9) shows that there is no correlation
between the intrinsic X-ray luminosity and the amount of absorption.

\section{Summary and Conclusions}

In this paper we have described the detailed analysis of the spatial,
temporal and spectral properties of the discrete X-ray sources detected
with a 72 ks {\it Chandra} ACIS-S observation of the
Antennae galaxies. Our results are summarized below:

\begin{enumerate}
\item
We used two different algorithms for source detections: the sliding
box {\it celldetect} and the wavelet transform {\it wavdetect}.
We find that both algorithms lead to the detection of 49 X-ray sources.
Three additional sources are found with {\it celldetect}, but visual
inspections of the image shows that they are peaks of the diffuse
emission of the Antennae. The measured net source counts from the two
detection methods typically agree within 30\%, the biggest discrepancies
occurring in confuse fields, where image inspection shows that
{\it wavdetect} performs better than {\it celldetect}. For this reason
we use the {\it wavdetect} results to derive the source fluxes.
\item
 We detect a total of 49 sources within the optical area of the Antennae.
The limiting luminosity is  $\sim10^{38}$\ergs~ ($\rm{H_{0}=50}$\kmsmpc;
$\sim5\times10^{37}$\ergs, for  $\rm{H_{0}=75}$\kmsmpc). 31 sources (38 for
$\rm{H_{0}=75}$\kmsmpc) have luminosities below $10^{39}$\ergs~ 
while 18 (11) have luminosities well in the ULX range, i.e.
well in excess of the Eddington limit for a neutron star accretor. Based on the expected number
of serendipitous sources in the field, only 2-3 sources may not be
associated with the Antennae, and the probability of chance
association is $<1$ for the ULXs (see Paper I).
\item
We analyzed the spatial extent of the 20 sources detected with 100 counts
or more, by comparing the spread of counts with the {\it Chandra}$+$ACIS-S
PSF. All but 6 sources are point-like. Six sources have extended 
components, with extent ranging between 3" and 15" ($\sim400 - 2100$pc).
These extended sources include the two nuclear regions.
\item
Two luminous sources are found to vary within the observation with the KS test.
By comparing the {\it Chandra} data with the ROSAT HRI observations
of the Antennae (5" angular resolution) of Fabbiano et al (1997),
we find evidence of long-term variability (year timescales) in 3 more sources, one of
which includes a \chandra variable source.
\item
We extracted ACIS-S spectra for all the 49 detected sources, and derived
X-ray colors for all of them. Model fitting was performed on all the 31
sources with more than $\sim50$ net counts and  for which the data could be grouped in at least 5 spectral
bins. 
\item
We fitted the data with both an absorbed power-law model and a thermal Raymond 
model with solar abundances and absorption. In both cases absorption columns in excess
of the Galactic line of sight $N_H$ are favored. Two-component models (power-law plus
Raymond) were fitted to
the spectra of 10 sources for which a large enough number of counts
was detected. 
 With the exception of one source, no evidence was found in these 
composite models for an intrinsically absorbed power-law.
Using more complex models, such as power-law plus multi-temperature black-body,
which fit well X-ray binary spectra, did not produce any detectable improvement in
the fit.
Three sources show significant fit residuals suggesting the presence of emission
lines. Six sources appear to have absorbed spectra, with \nh~ higher
than 10 times the Galactic along the line of sight.

\item
We used Hardness Ratio (HR) diagrams, to derive spectral information from all 
the detected sources. We experimented with different selections of energy boundaries,
and chose 0.3-1~keV (Soft), 1-2.5~keV (Medium), and 2.5-7.0~keV (Hard), because
these offer the best differentiation of models in an HR-HR diagram. We calibrated 
these diagrams by simulating both power-law and thermal spectra for a variety 
of parameters, and compared these results with the results of the spectral fits,
with good agreement. By using HR diagrams, we find that the majority of 
low-luminosity sources have soft spectra.
\item
By plotting the power-law photon indices from the single power-law fits 
against source luminosity, we find an indication of an overall spectral
trend, with the most luminous sources exhibiting very hard spectra ($\Gamma
\sim 1.2$), while softer emission is prevalent at the low luminosities
(with values of $\Gamma > 3$ in some cases).
No luminosity -- $N_H$ trend is observed.

\end{enumerate}

Our results suggest that a good fraction 
of the most luminous sources (ULXs) in the Antennae may be
compact accretion binaries. This conclusion is 
supported by the presence of variability in  few sources and by the
luminosity -- photon index  trend suggesting
that hard sources (consistent with X-ray binary emission)
dominate the source population at high luminosities.
At lower luminosities, instead, the softer X-ray spectra
may suggest the presence of supernova remnant emission
as well. We must however remember that these fainter
spectra may include a sizeable amount of residual emission from
the soft hot ISM in the spectral extraction region
(see Paper~I).

A detailed discussion of these results, augmented by
the X-ray analysis of the average spectral properties
in different source luminosity ranges, and supported
by multi-wavelength comparison to constrain the nature
of the emitting sources, is presented in Paper-III (Zezas \etal  2002).

\acknowledgments

 We thank the CXC DS and SDS teams for their efforts in reducing the data and 
developing the software used for the reduction (SDP) and analysis
(CIAO). We thank Martin Ward, Jeff McClintock, Andrea Prestwich and Phil Kaaret for
useful discussions on these results. 
This work was supported by NASA contract NAS~8--39073 (CXC)
and NAS8-38248 (HRC).

\clearpage

\input{tables.tex}

\clearpage


\begin{figure}
\caption{ Images of the Antennae in the full
(0.3-10.0keV), soft (0.3-2.0keV), medium (2.0-4.0keV) and hard
(4.0-10.0keV) (clockwise) bands. In the raw full band image we present both the
celldetect (red) and wavdetect (white) $3\sigma$ source ellipses, whereas in the other three images
(which are adaptivelly smoothed) we present only the sources detected
by wavdetect. The numbering convention is the same as in Table~1.}
\end{figure}

\begin{figure}
\caption{ Examples for fits of the radial profiles of point-like
(first row) and extended sources (second and third row)  with
their corresponding model PSFs. The bottom panel shows the fit residuals}.
\end{figure}

\begin{figure}
\caption{ The cumulative distributions of the photon arrival times (0.3-7keV)
for sources 14 (left) and 44 (right) compared with the distribution of
the background events (dashed line).}
\end{figure}

\begin{figure}
\caption{ The source (left) and background (right) regions, used to
extract the spectra for each source, overlaid on a raw full band
(0.3-10.0~keV) image. The two images are on the same scale and for
clarity we give the source numbers (as in Table~1) only in the left image. In the right
image the red circles are the background regions and the white circles
are the regions excluded from the background.}
\end{figure}

\begin{figure}
\caption{The spectra of the individual sources in the Antennae
galaxies together with the best fit power-law model and N$\rm{_H}$ -
$\Gamma$ confidence contours.  The
contours are at the 1$\sigma$ (dotted line), 2$\sigma$ (dashed line)
and 3$\sigma$ (solid line) confidence levels for two
interesting parameters. The Galactic line-of-sight column density is
noted by a horizontal dashed line. 
The bottom panel
presents the ratio of the model and  the data.  }
\end{figure}

\begin{figure}
\caption{ The spectra of the sources with detected emission lines
together with the best fit model. The bottom panel shows the fit residuals.}
\end{figure}

\begin{figure}
\caption{ Hardness ratio diagrams for the bright discrete sources in the
Antennae. The red grids correspond to power-law models and the green
grids correspond to Raymond-Smith models. The arrows point towards
increasing values of each parameter. In both models the dashed lines
correspond to N$\rm{_H}$ of (0.01, 0.025,  0.05, 0.075, 0.10, 0.12,
0.25, 0.5)$\times10^{22}$\cmsq  }
\end{figure}

\begin{figure}
\caption{ Hardness ratio diagrams for the faint discrete sources in the
Antennae. The model grids are the same as in figure 7.}
\end{figure}

\begin{figure}
\caption{ Plots of the luminosity against the  photon index (left) and
column density (right)    for point-like (red), and extended (blue)
sources. The three black points with the error-bars in the bottom of the
$\rm{L_{X} - \Gamma}$ plot show the mean and the standard deviation of
the photon indices in each luminosity range (in \ergs). In the $\rm{L_{X} -
N_{H}}$ plot the dotted blue line marks 
the Galactin line-of-sight column density.}
\end{figure}

\eject


\setcounter{figure}{0}

\begin{figure}
\caption{}
\end{figure}				                
					                
\begin{figure}				                
\rotatebox{360}{\includegraphics[height=5.0cm]{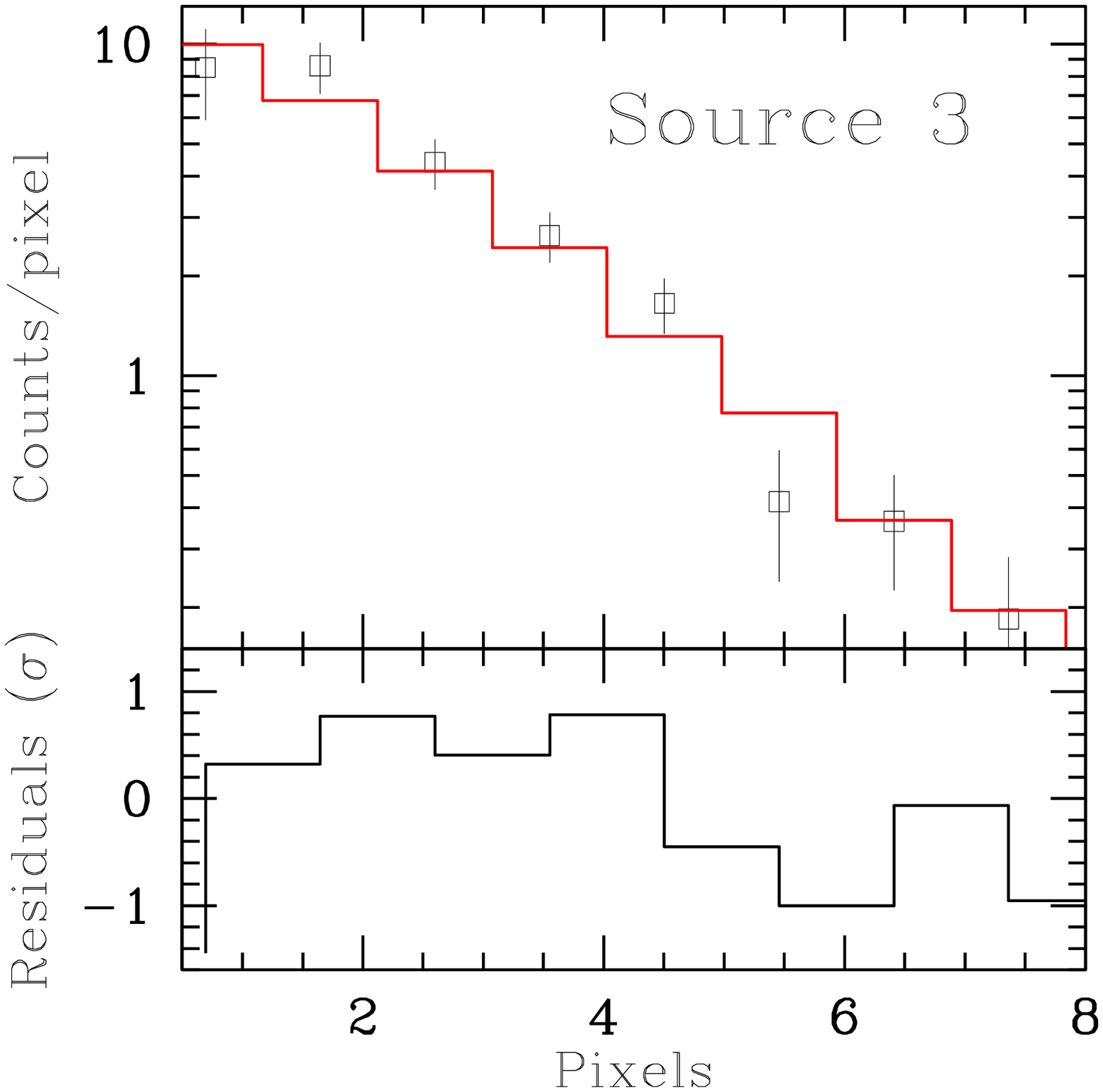}} 
\rotatebox{360}{\includegraphics[height=5.0cm]{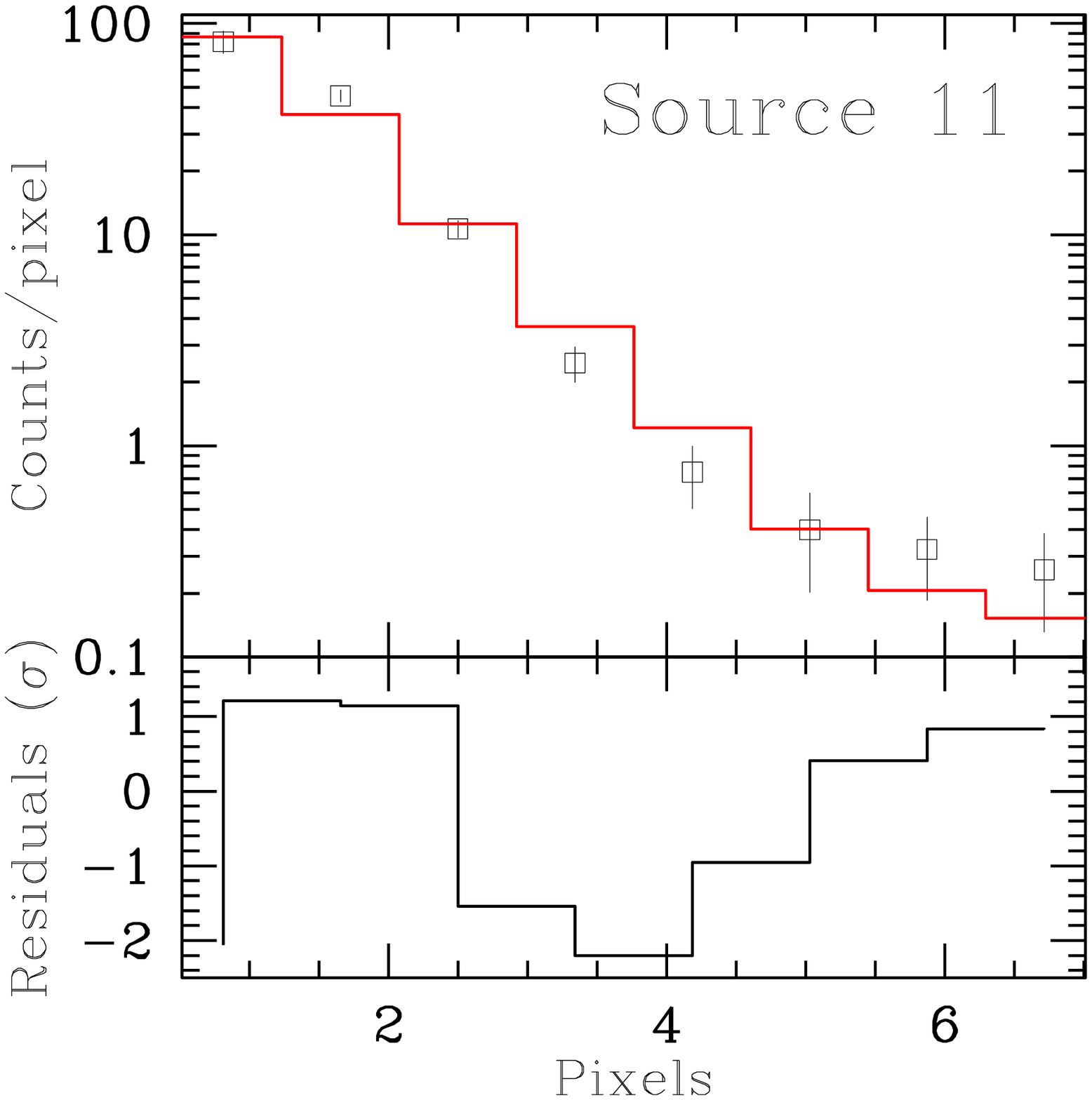}} 
\rotatebox{360}{\includegraphics[height=5.0cm]{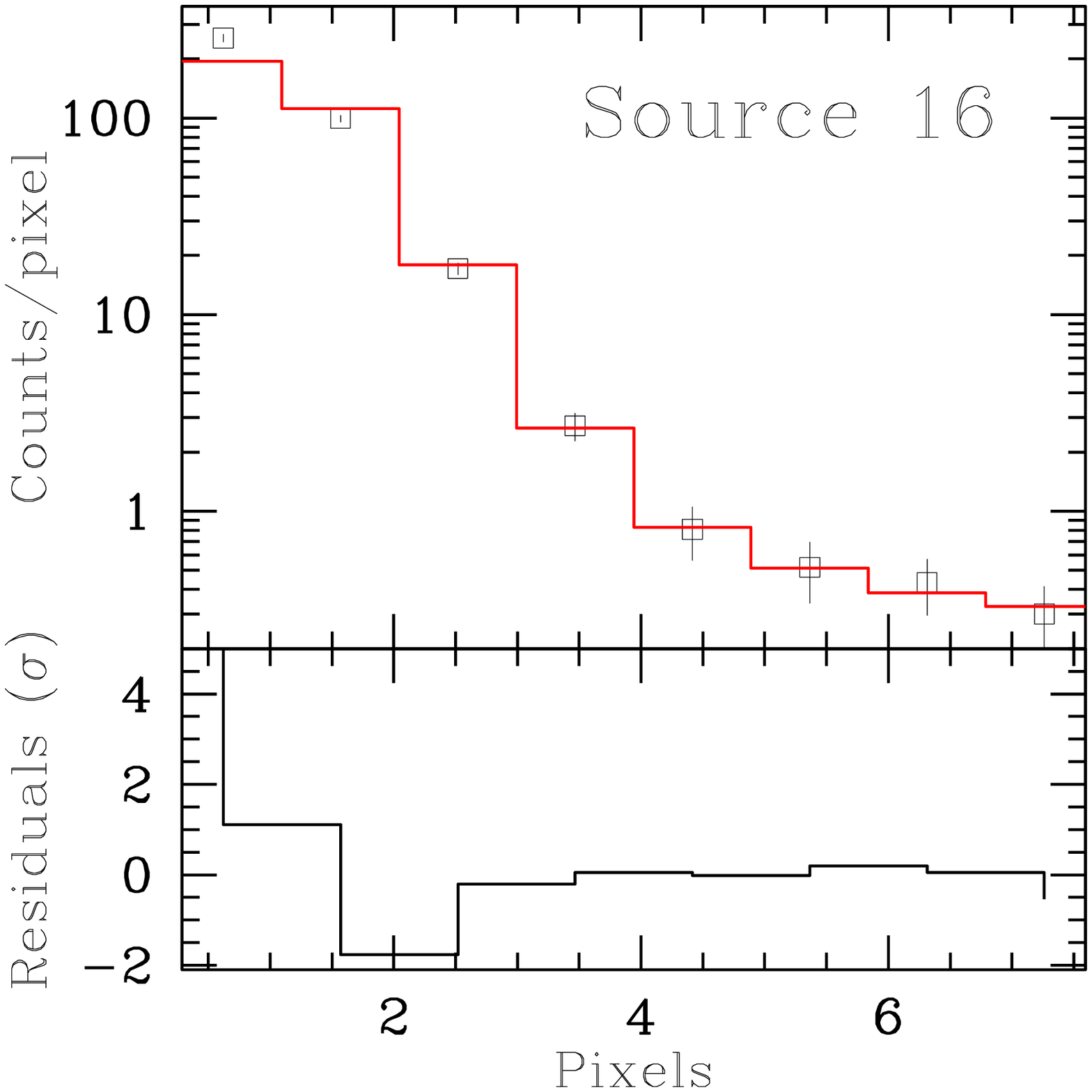}} 
\rotatebox{360}{\includegraphics[height=5.0cm]{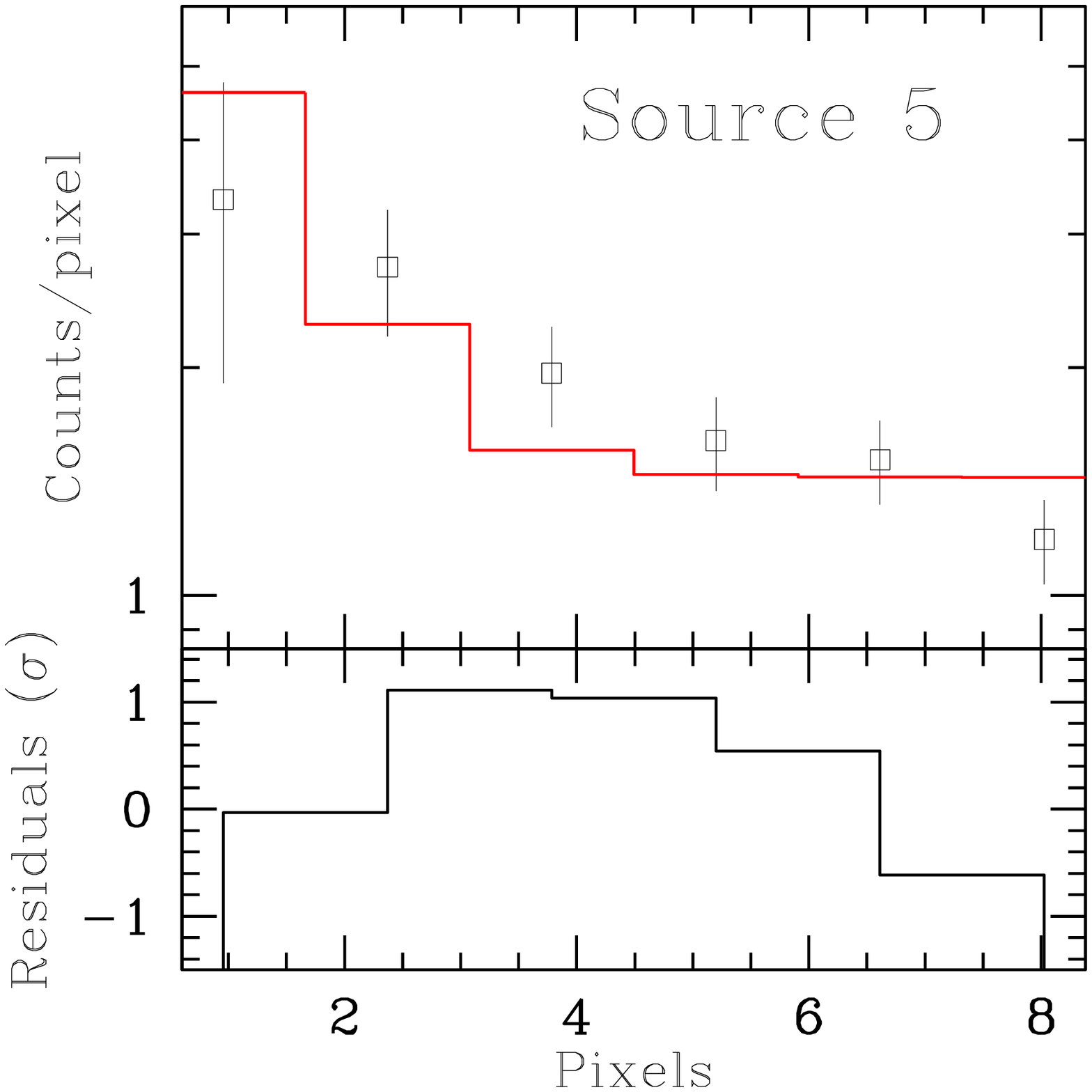}} 
\rotatebox{360}{\includegraphics[height=5.0cm]{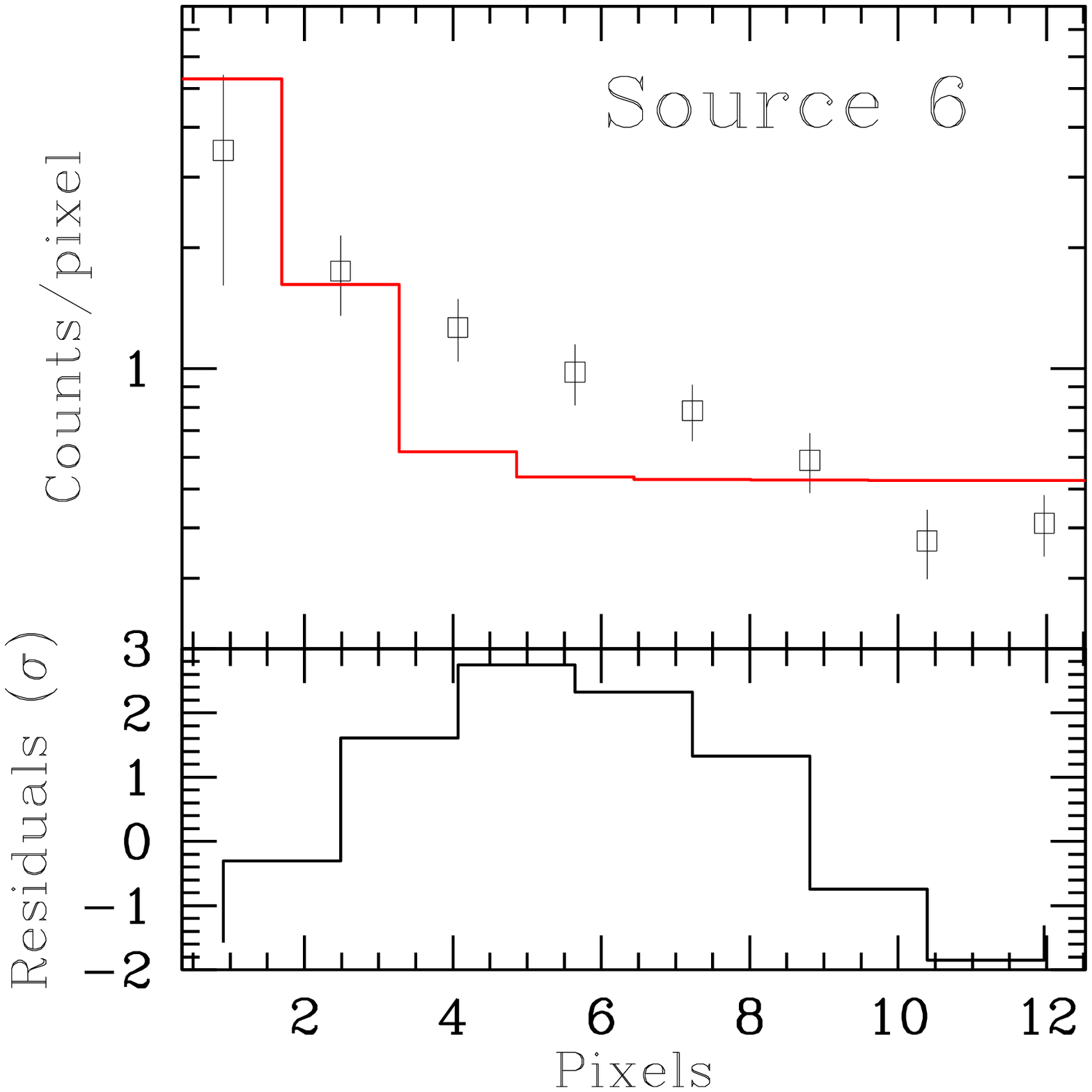}} 
\rotatebox{360}{\includegraphics[height=5.0cm]{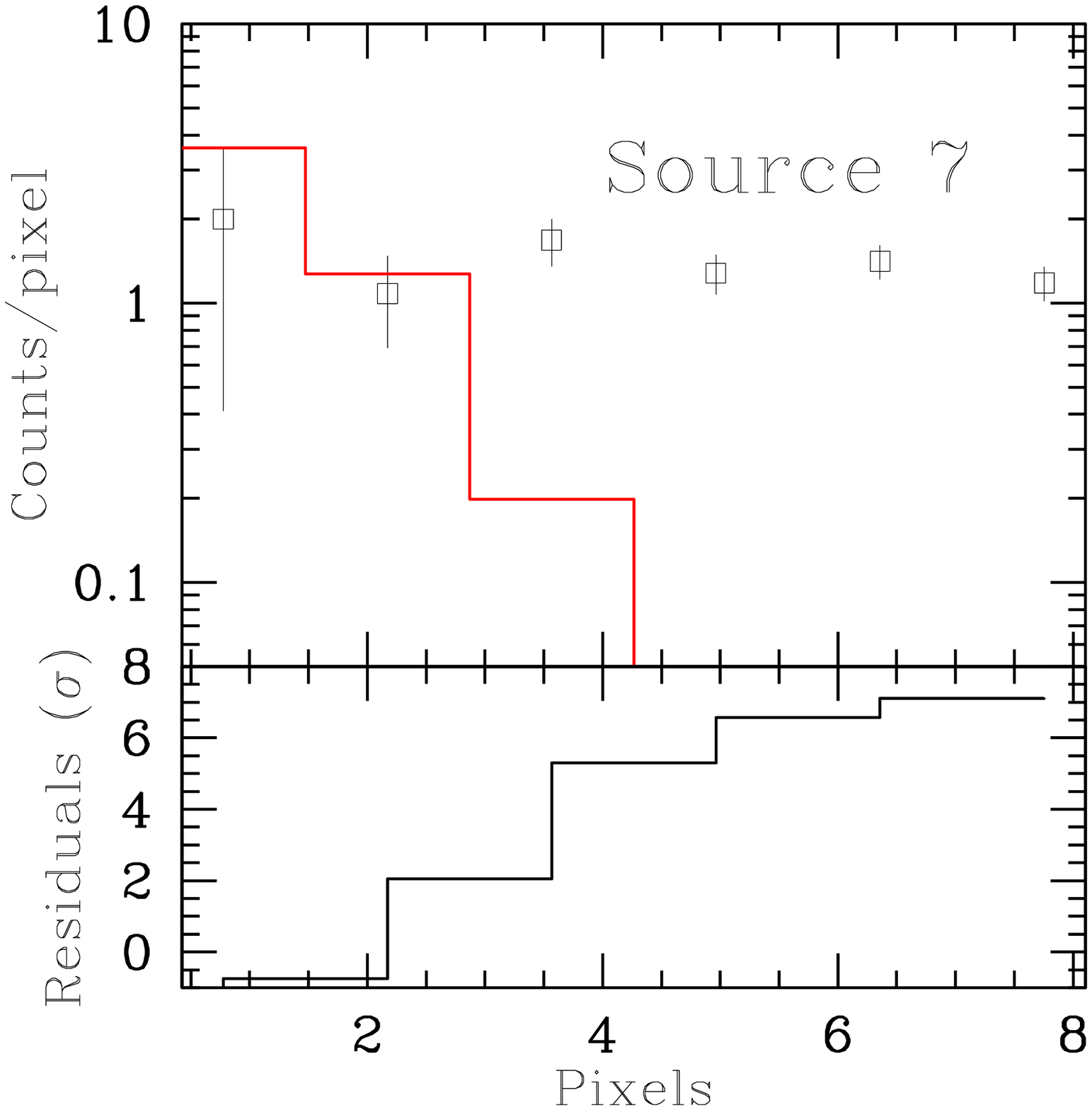}} 
\rotatebox{360}{\includegraphics[height=5.0cm]{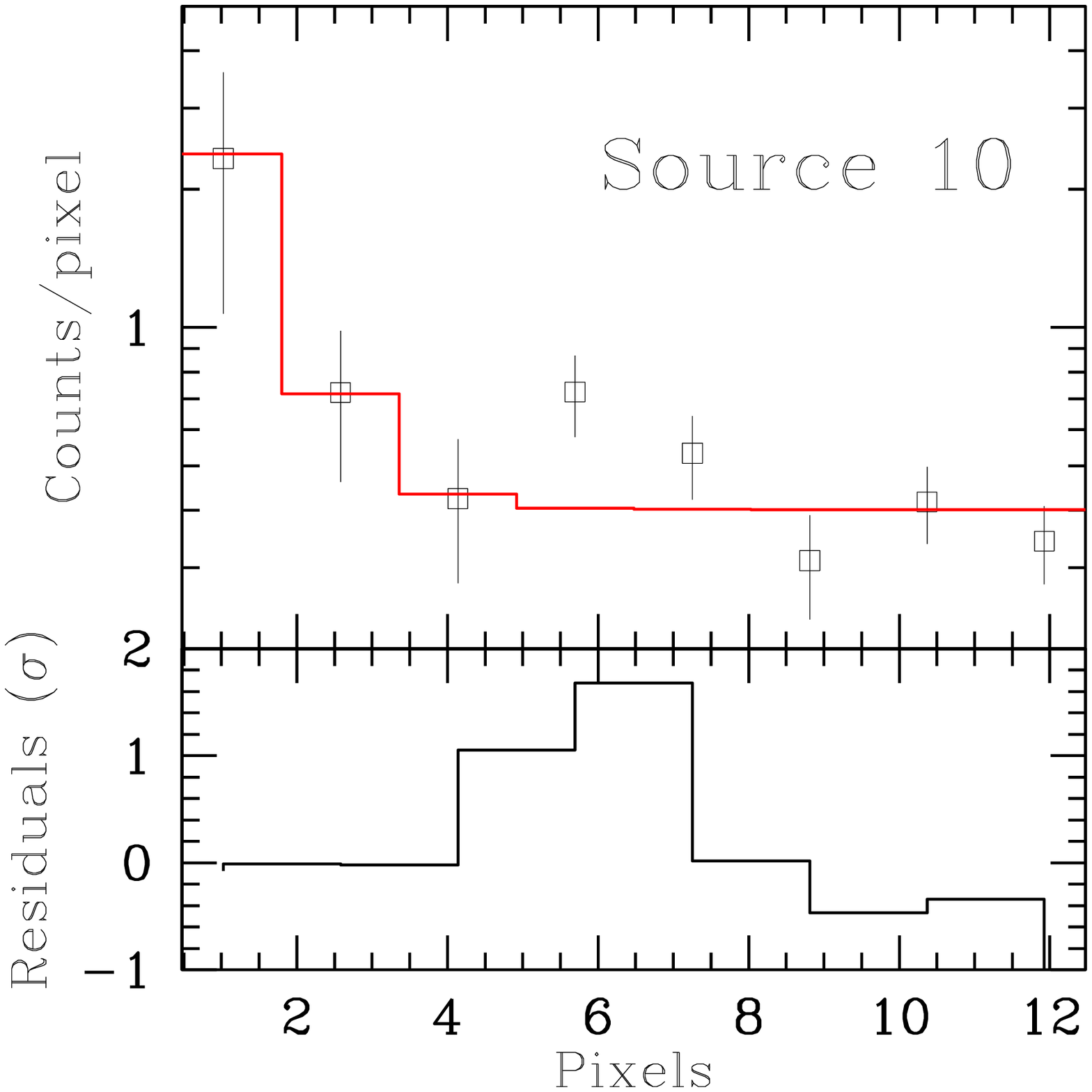}} 
\rotatebox{360}{\includegraphics[height=5.0cm]{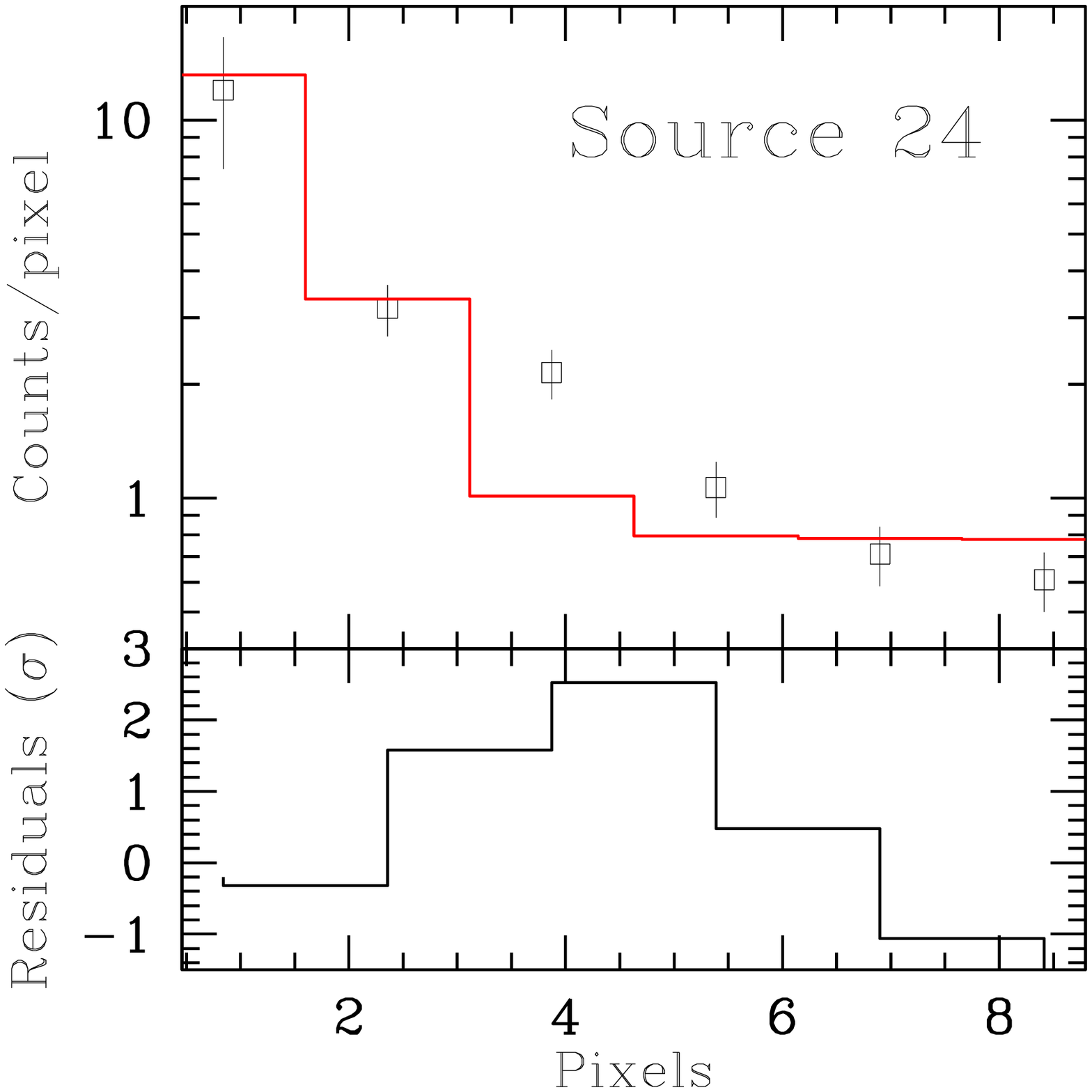}} 
\rotatebox{360}{\includegraphics[height=5.0cm]{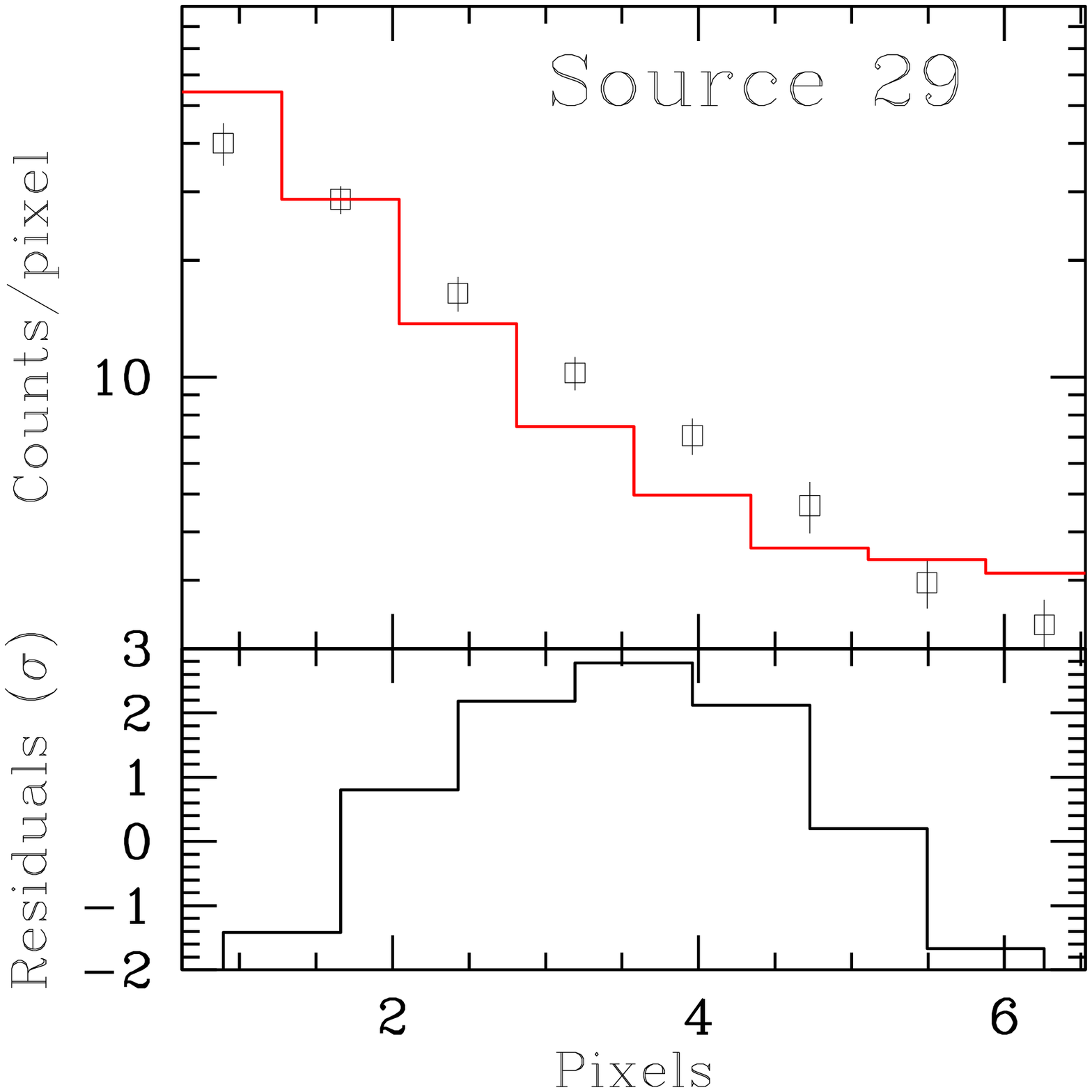}} 
\caption{}
\end{figure}				                

\eject
					                
\begin{figure}				                
\rotatebox{270}{\includegraphics[height=7.0cm]{f3a.eps}} 
\rotatebox{270}{\includegraphics[height=7.0cm]{f3b.eps}}
\caption{}
\end{figure}

\begin{figure}				                
\rotatebox{360}{\includegraphics[height=9.0cm]{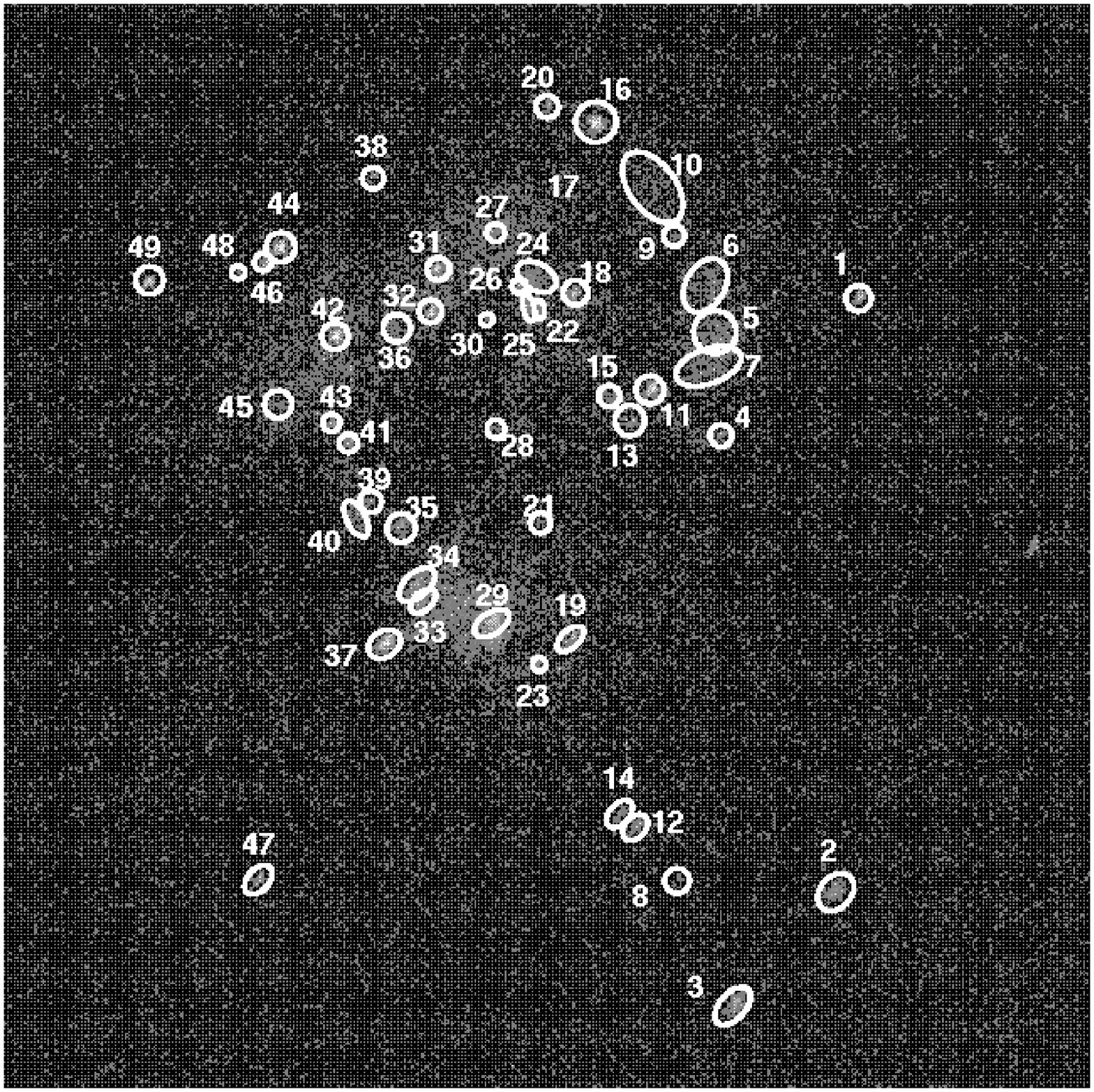}} 
\rotatebox{360}{\includegraphics[height=9.0cm]{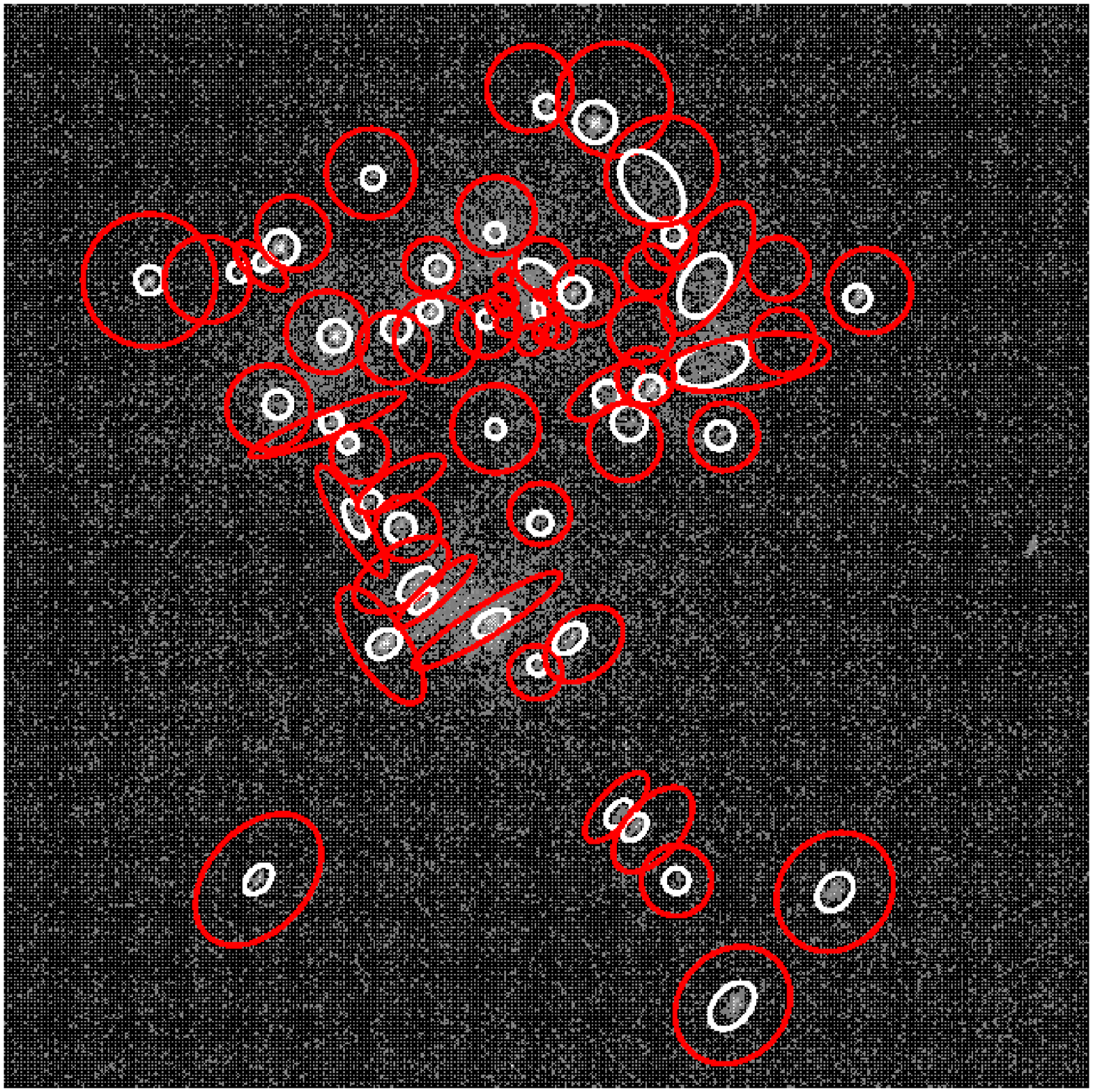}} 
\caption{}
\end{figure}				                

\clearpage	
				                
\begin{figure}				                
\rotatebox{270}{\includegraphics[width=11.0cm]{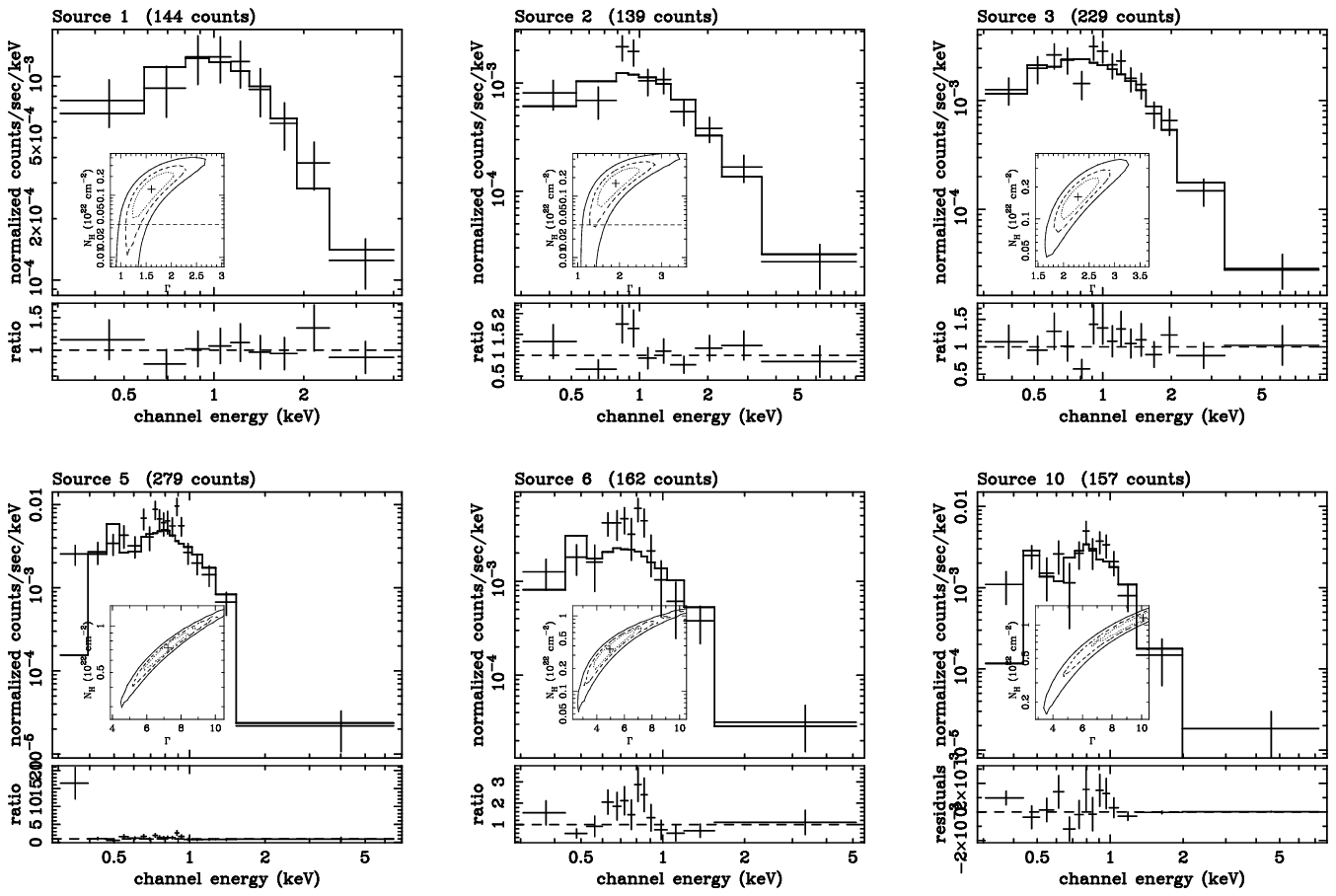}} 
\rotatebox{270}{\includegraphics[width=11.0cm]{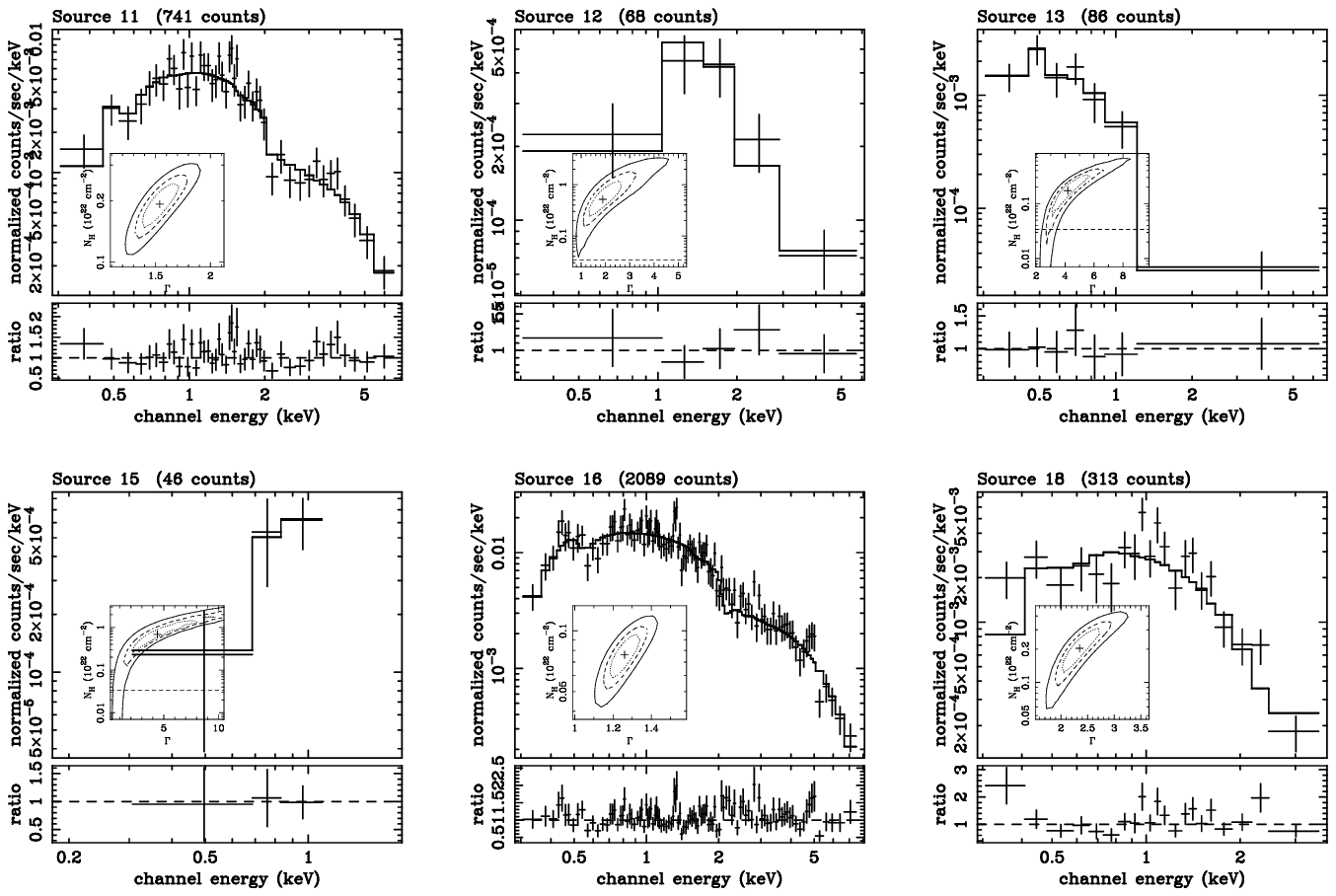}} 
\end{figure}					                
\newpage
\begin{figure}					                
\rotatebox{270}{\includegraphics[width=11.0cm]{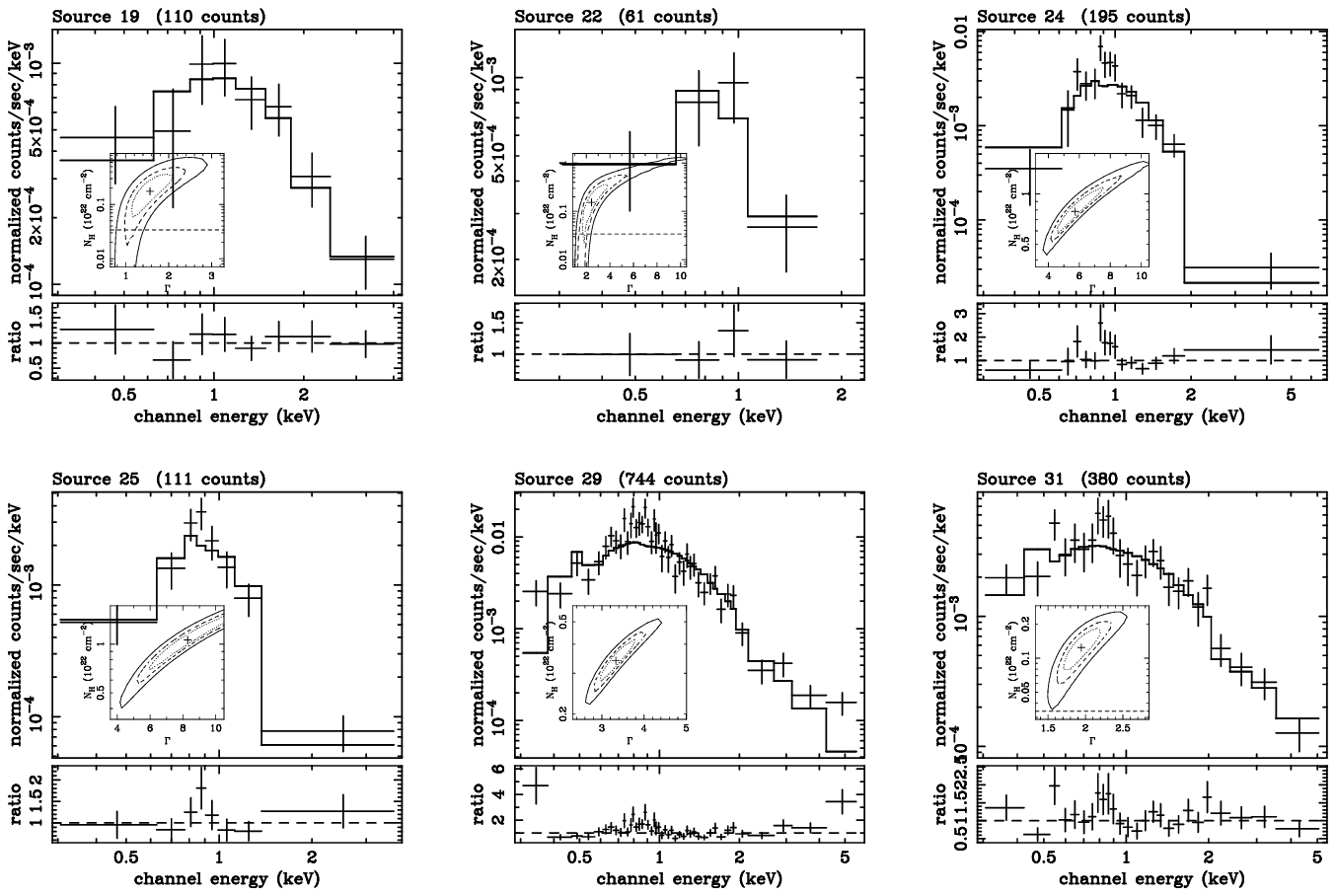}} 
\rotatebox{270}{\includegraphics[width=11.0cm]{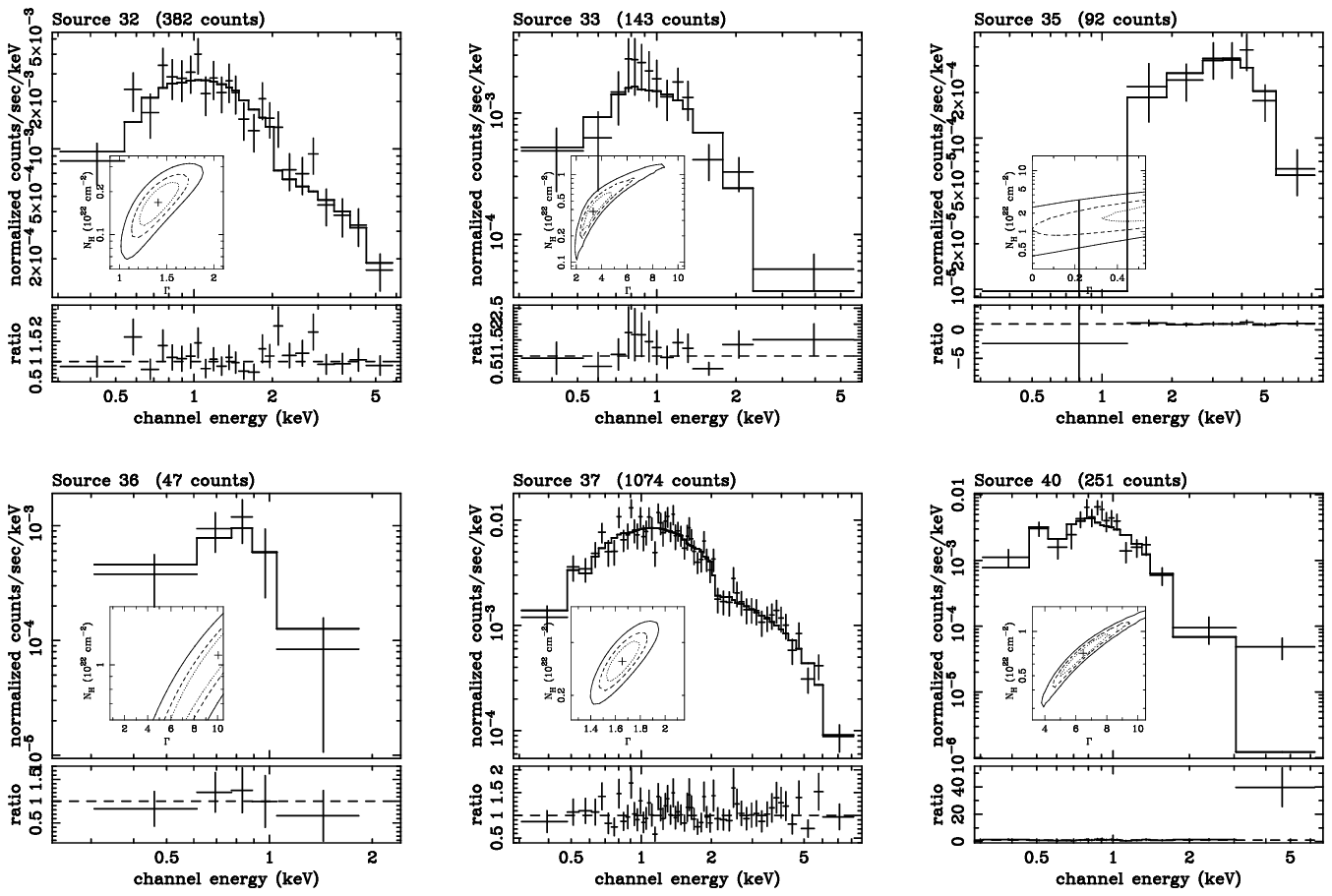}} 
\end{figure}					                
\newpage
\begin{figure}					                
\rotatebox{270}{\includegraphics[width=11.0cm]{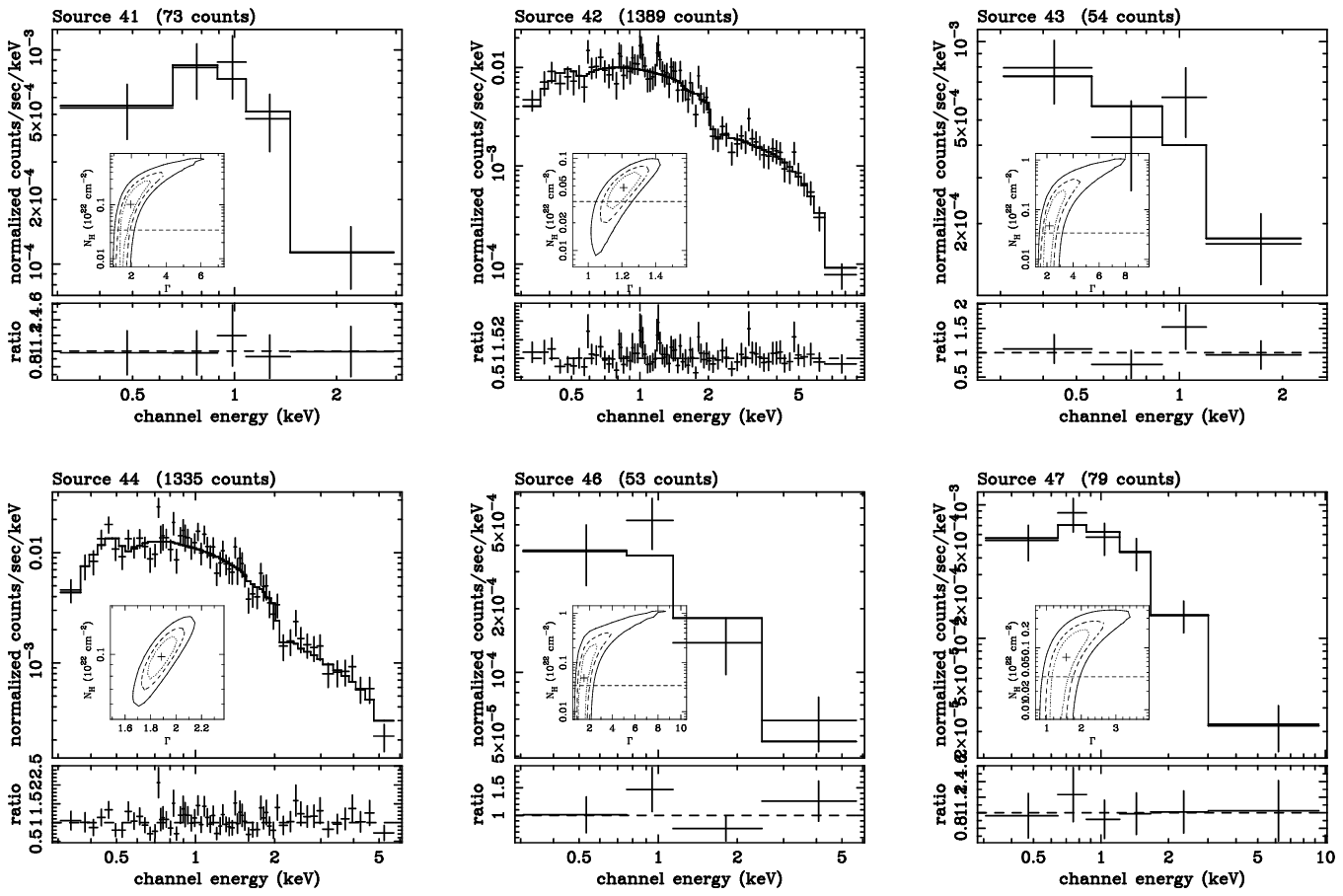}} 
\rotatebox{270}{\includegraphics[width=5.25cm]{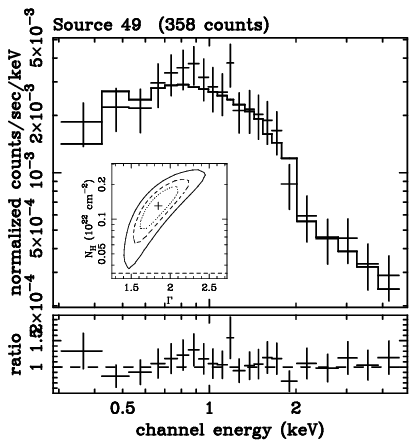}} 
\caption{}
\end{figure}				                

\clearpage
					                
\begin{figure}				                
\rotatebox{270}{\includegraphics[width=5.5cm]{f6a.eps}} 
\rotatebox{270}{\includegraphics[width=5.5cm]{f6b.eps}} 
\rotatebox{270}{\includegraphics[width=5.5cm]{f6c.eps}} 
\caption{}
\end{figure}				                
					                
\begin{figure}				                
{\includegraphics[width=9.0cm]{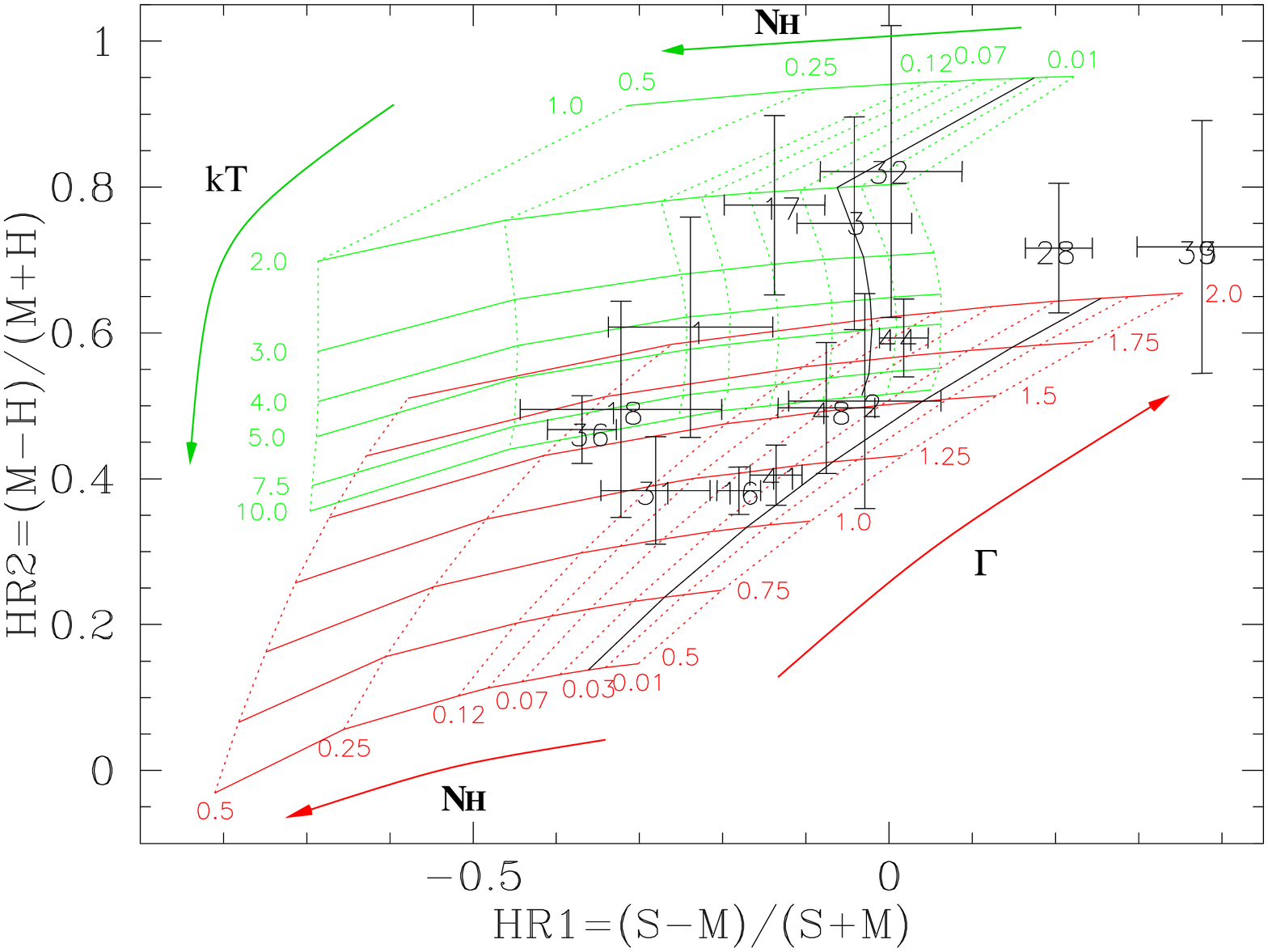}} 
{\includegraphics[width=9.0cm]{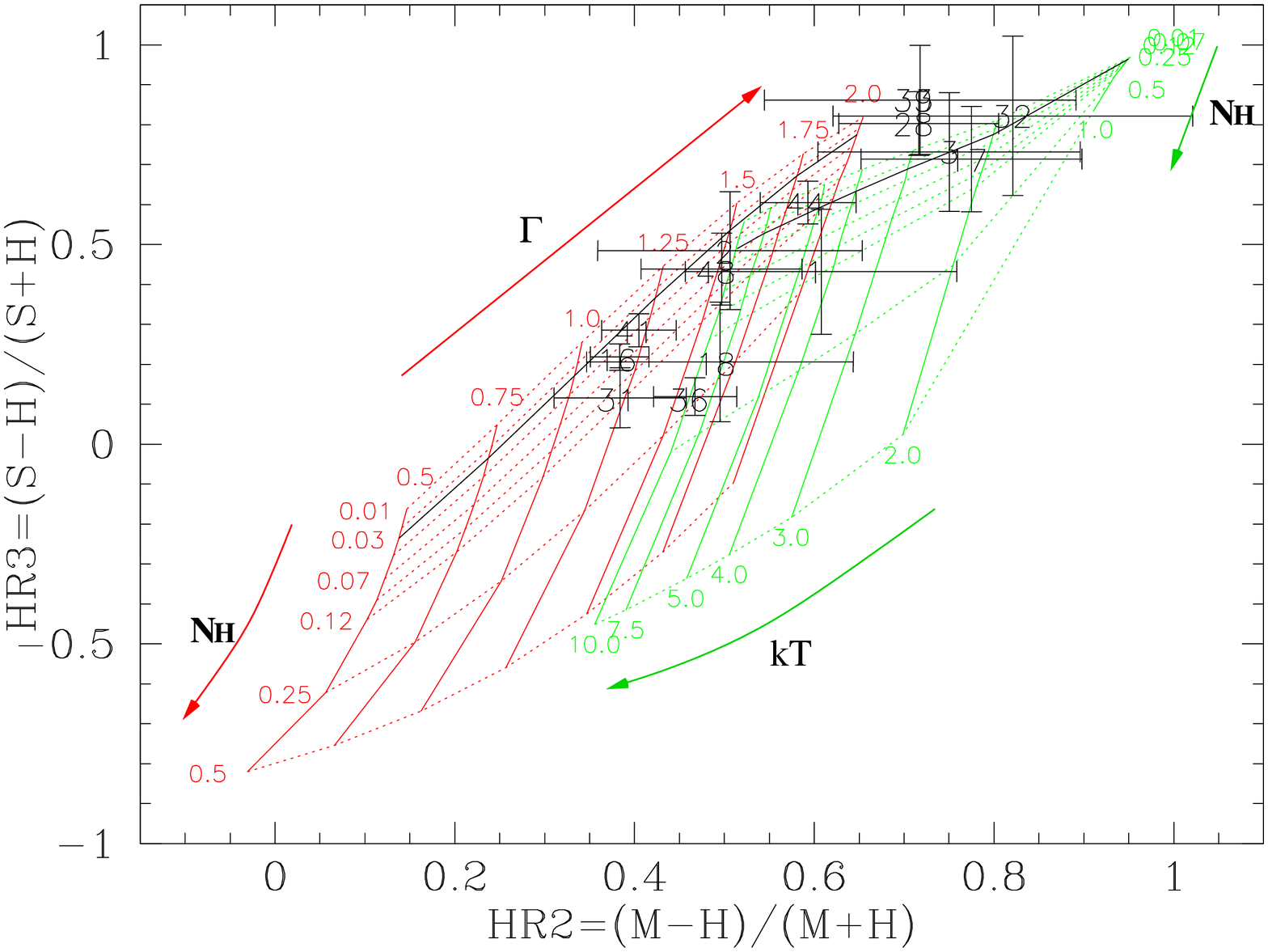}} 
{\includegraphics[width=9.0cm]{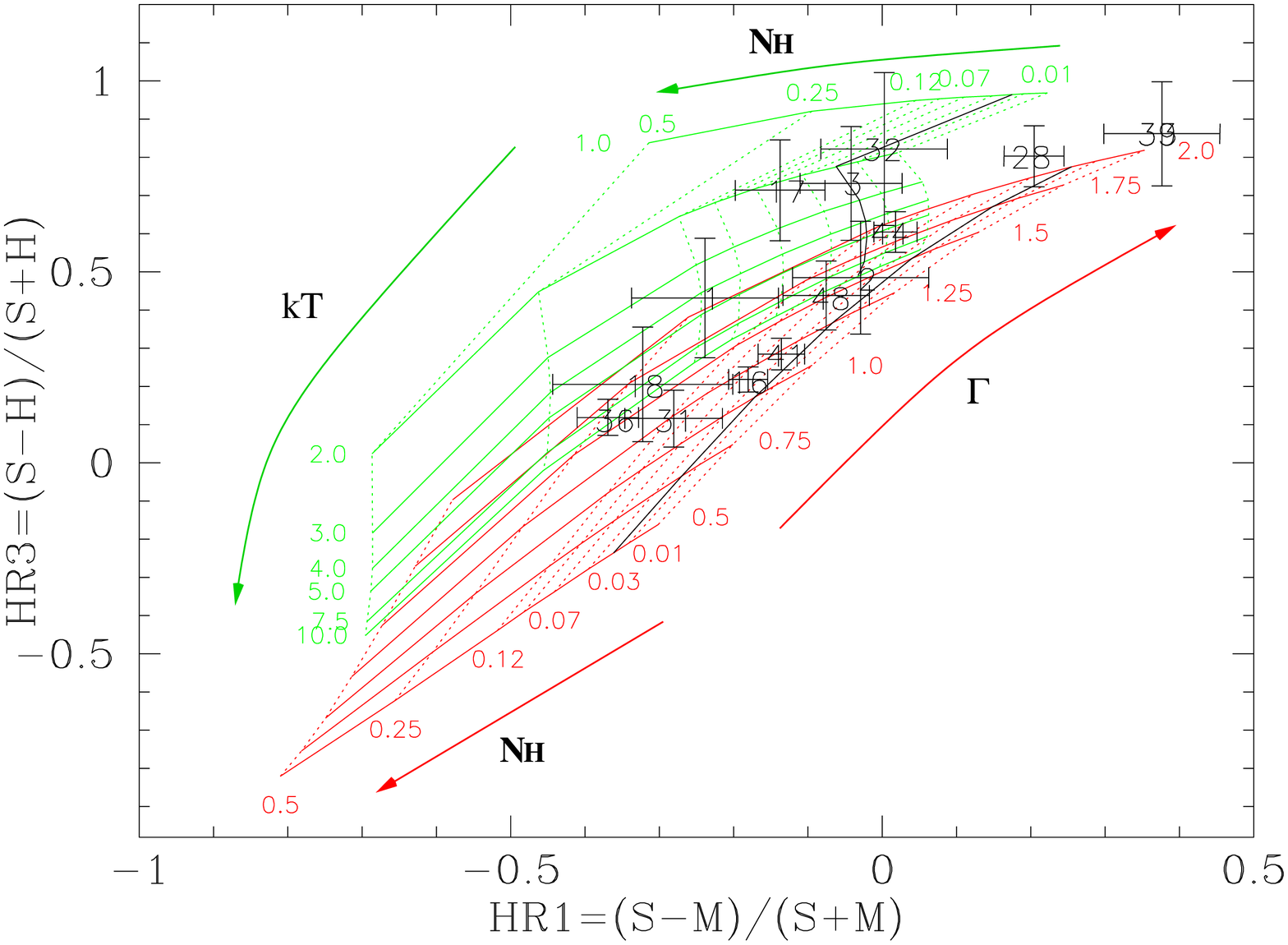}} 
\caption{}		      
\end{figure}		                
			                
\begin{figure}		                
{\includegraphics[width=9.0cm]{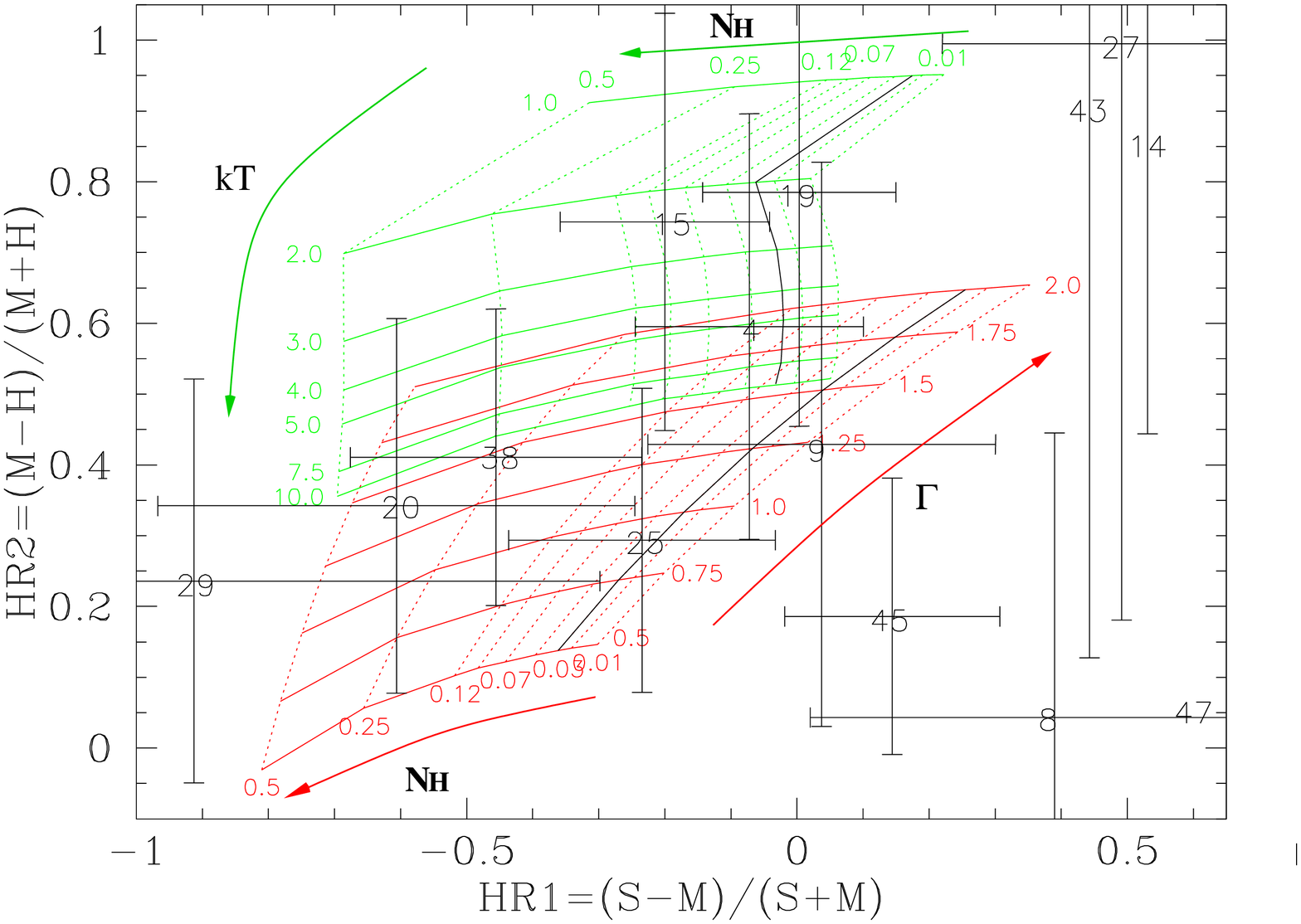}} 
{\includegraphics[width=9.0cm]{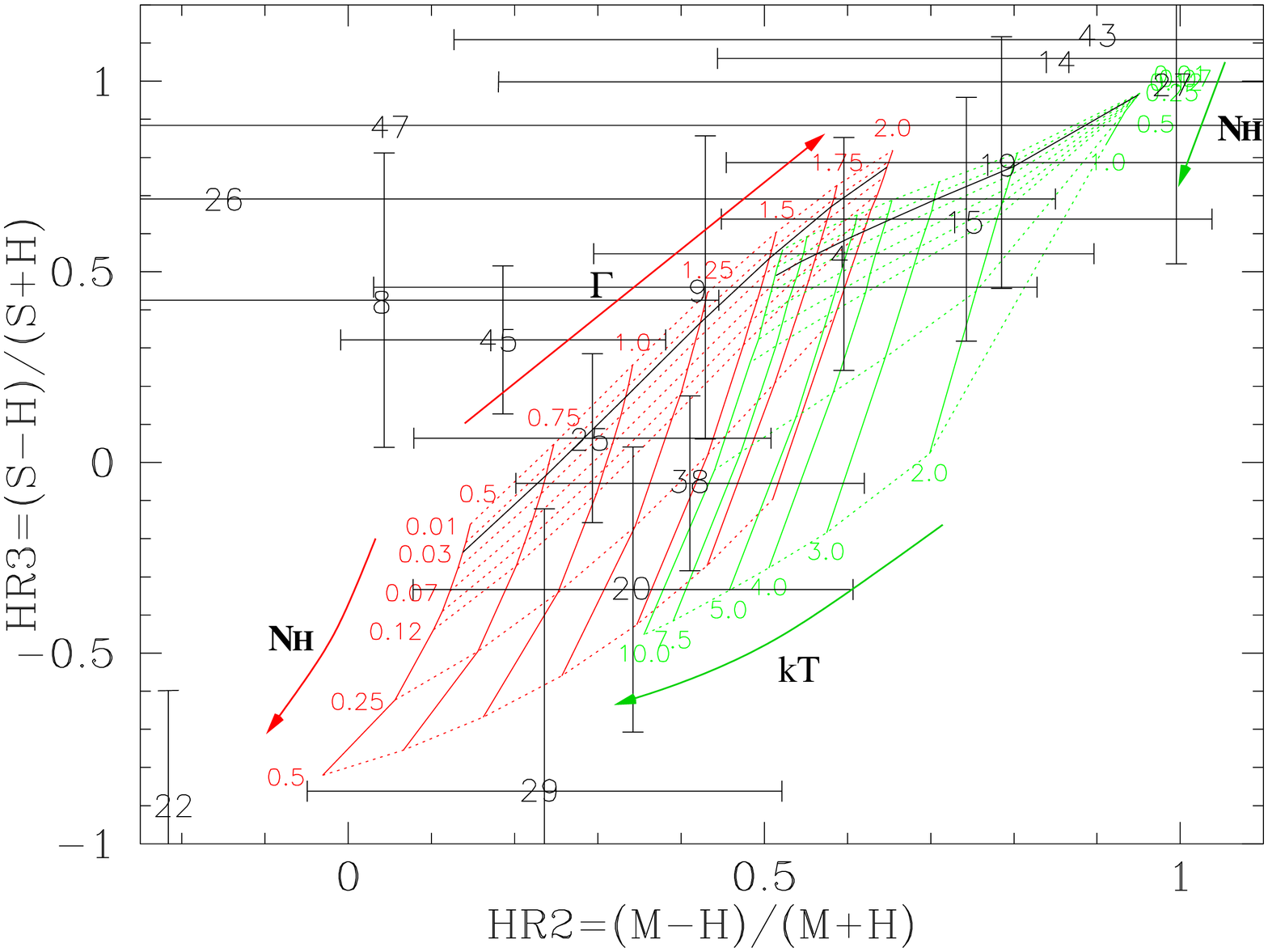}} 
{\includegraphics[width=9.0cm]{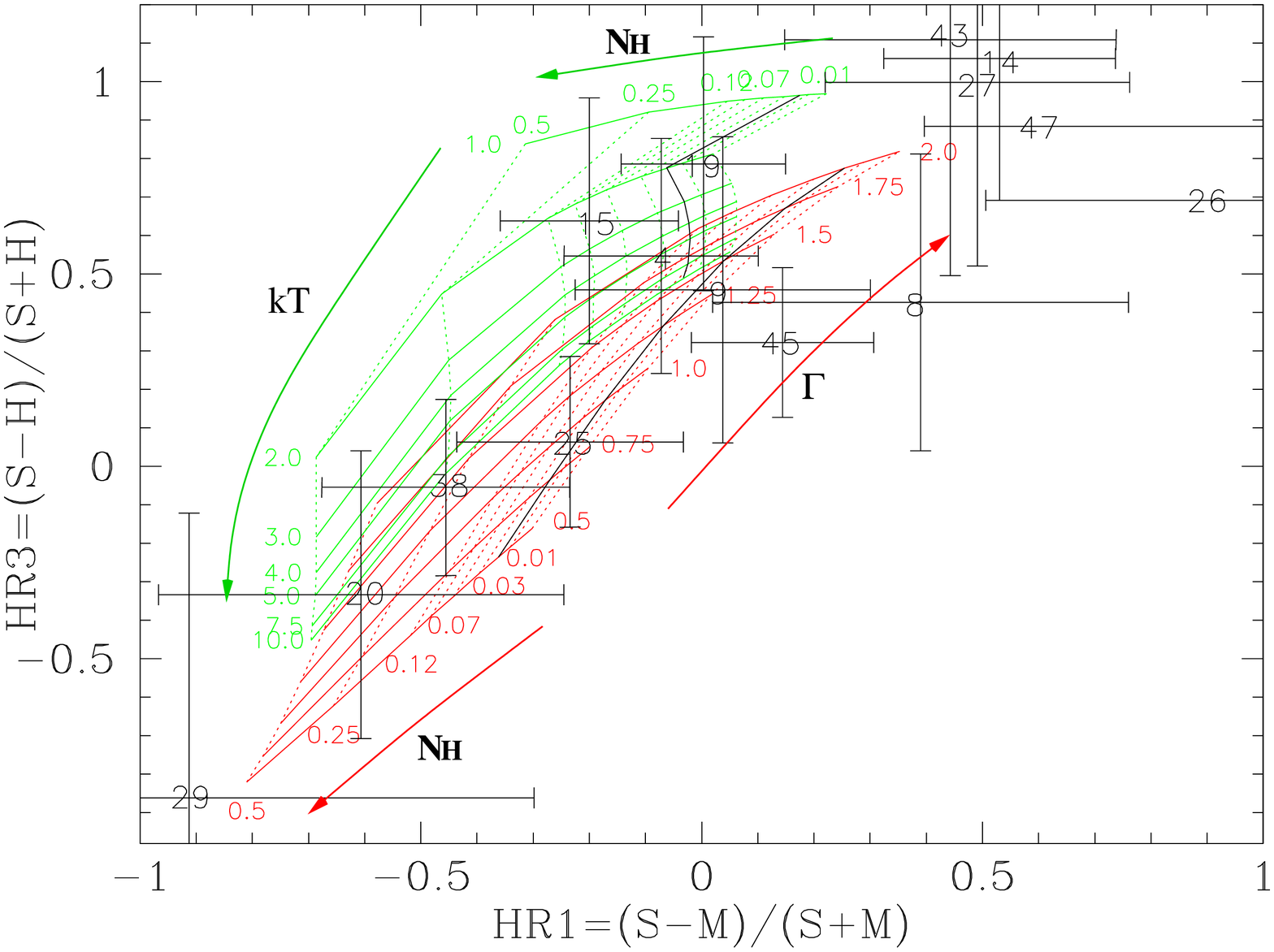}} 
\caption{}
\end{figure}

\begin{figure}				                
\rotatebox{270}{\includegraphics[height=9.0cm]{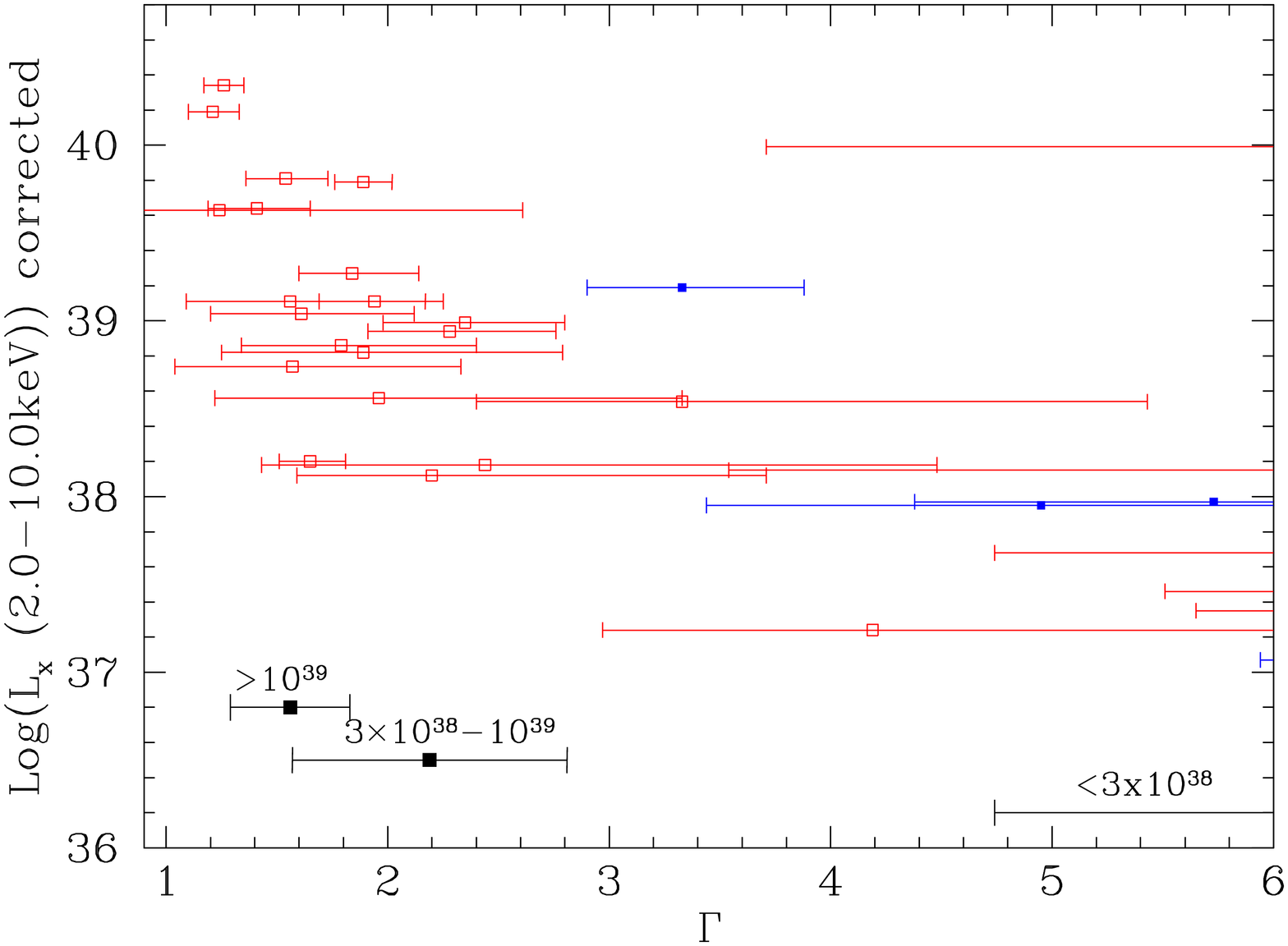}}
\rotatebox{270}{\includegraphics[height=9.0cm]{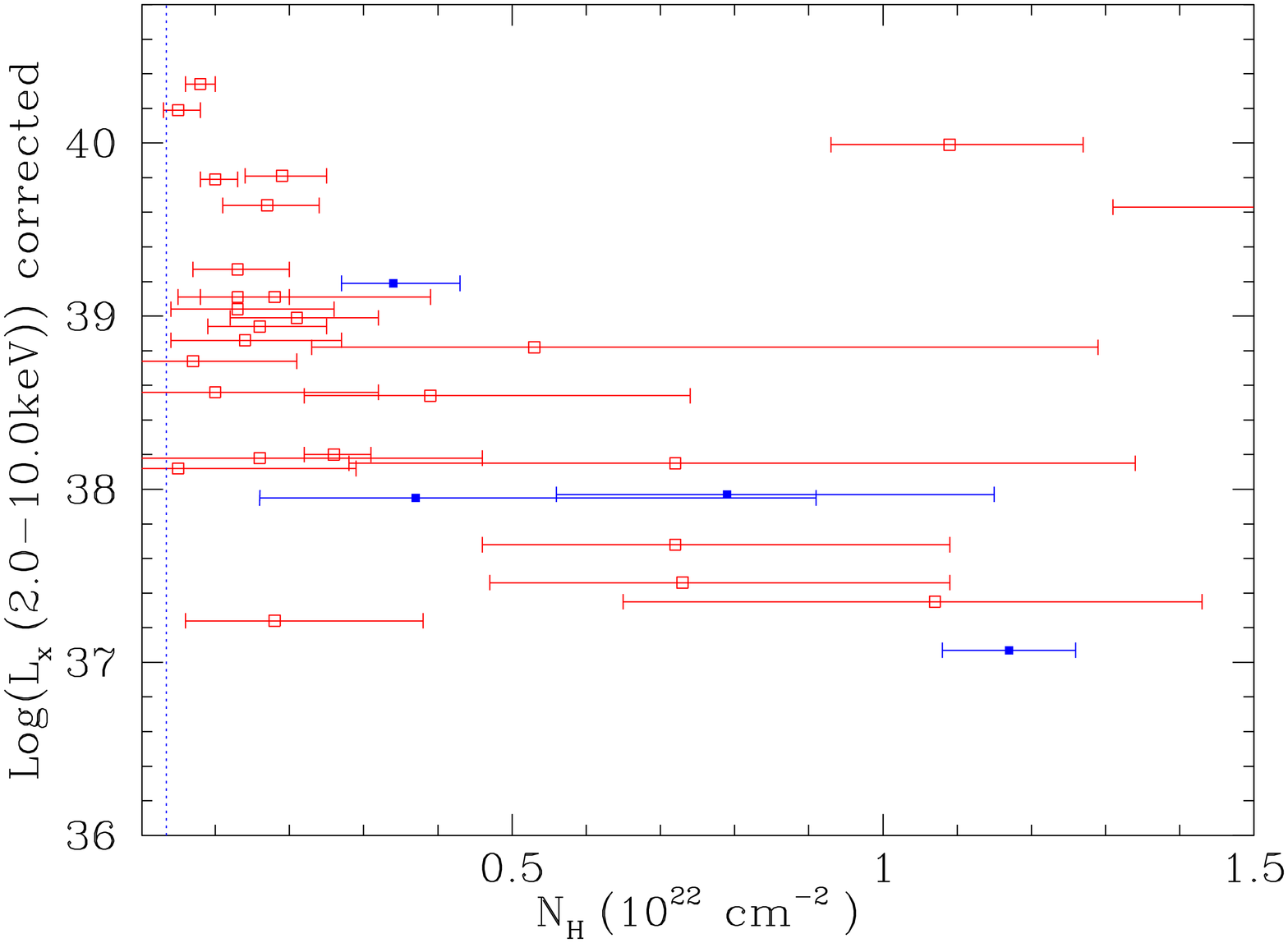}}
\caption{}
\end{figure}

\end{document}

%% file: tables.tex


\makeatletter
\def\jnl@aj{AJ}
\ifx\revtex@jnl\jnl@aj\let\tablebreak=\nl\fi
\makeatother
\ptlandscape
\begin{deluxetable}{lcccccccccccccccc}
\tabletypesize{\scriptsize}
\rotate
\tablecolumns{17}
\tablewidth{0pt}
\tablecaption{ Source list (0.3-10.0keV)}
\tablehead{
\colhead{}    &  \colhead{}    &  \multicolumn{6}{c}{Wavelets} &  
\multicolumn{6}{c}{Celldetect}  &  \colhead{}  \\ 
\cline{3-8} \cline{9-14} \\
 \colhead{Src } &\colhead{CXO name }	&\colhead{RA  } &\colhead{Dec
} &\colhead{net counts }
&\colhead{Bkg} &\colhead{S/N}  &\colhead{ext$^1$}&\colhead{RA  } &\colhead{Dec
} &\colhead{net counts }
&\colhead{Bkg } &\colhead{S/N} &\colhead{ext$^1$} &\colhead{Log($\rm{L_X^{obs}}$)$^{2}$}&\colhead{Log($\rm{L_X^{corr}}$)$^{2}$}\\
\colhead{ } &\colhead{CXOANT }	&\colhead{ J2000 } &\colhead{
J2000 } &\colhead{$\pm$error  }
&\colhead{ } &\colhead{($\sigma$) }  &\colhead{ }  &\colhead{ J2000 } &\colhead{
J2000 } &\colhead{$\pm$error  }
&\colhead{ } &\colhead{($\sigma$) } &\colhead{ }&\multicolumn{2}{c}{(0.1-10.0) keV } \\
\colhead{ (1) }	&\colhead{ (2)}
&\colhead{(3)} &\colhead{(4)} &\colhead{(5)} &\colhead{(6)}
&\colhead{(7)} &\colhead{(8)} &\colhead{(9)} & \colhead{(10)}
&\colhead{(11)} &\colhead{(12)} &\colhead{(13)}&\colhead{(14)} &\colhead{(15)}&\colhead{(16)}} 
\startdata
1	&J120148.5-185207.5   &	12:01:48.47	&	-18:52:07.50	& $	142.3	\pm	12.3	$ &	8.7	&	34.9	&	1.0	&	12:01:48.47	&	-18:52:07.50	& $	135.3	\pm	13.8	$ &	7.7	&			9.8	&	0.9	&39.07   &       39.14	\\ 
2	&J120148.8-185403.6   &	12:01:48.83	&	-18:54:03.60	& $	148.9	\pm	13.3	$ &	27.1	&	23.7	&	1.0	&	12:01:48.82	&	-18:54:03.67	& $	148.0	\pm	15.5	$ &	25.0	&			9.6	&	1.0	&39.09   &       39.16	\\ 
3	&J120150.2-185425.7   &	12:01:50.19	&	-18:54:25.67	& $	233.1	\pm	16.0	$ &	23.9	&	39.1	&	0.8	&	12:01:50.19	&	-18:54:25.74	& $	243.2	\pm	18.1	$ &	19.8	&			13.4	&	0.9	&39.29   &       39.35	\\ 
4	&J120150.4-185234.5   &	12:01:50.38	&	-18:52:34.46	& $	47.0	\pm	7.6	$ &	11.0	&	10.6	&	1.2	&	12:01:50.38	&	-18:52:34.50	& $	39.7	\pm	9.6	$ &	10.3	&			4.1	&	1.0	&38.59   &       38.66	\\ 
5	&J120150.5-185215.8   &	12:01:50.47	&	-18:52:15.89	& $	66.3	\pm	8.9	$ &	13.7	&	13.8	&	1.5	&		&		& $				$ &		&				&				&38.74   &       38.81	\\ 
6	&J120150.5-185204.7   &	12:01:50.51	&	-18:52:04.77	& $	45.8	\pm	7.6	$ &	12.2	&	9.9	&	1.5	&	12:01:50.51	&	-18:52:05.30	& $	42.0	\pm	11.9	$ &	27.0	&			3.5	&	1.6	&38.58   &       38.65	\\ 
7	&J120150.6-185220.3   &	12:01:50.64	&	-18:52:20.32	& $	29.0	\pm	6.2	$ &	9.0	&	7.1	&	1.0	&		&		& $				$ &		&				&				&38.38   &       38.45	\\ 
8	&J120150.9-185401.9   &	12:01:50.97	&	-18:54:01.98	& $	18.9	\pm	5.3	$ &	9.1	&	4.5	&	0.9	&		&		& $				$ &		&				&				&38.20   &       38.26	\\
\enddata
\tablenotetext{1}{ Source extent in units of the PSF size.}
\tablenotetext{2}{Log of the observed and absorption corrected luminosity in
units of \ergs.}
\end{deluxetable}


\makeatletter
\def\jnl@aj{AJ}
\ifx\revtex@jnl\jnl@aj\let\tablebreak=\nl\fi
\makeatother
\ptlandscape
\begin{deluxetable}{lcccccccccccc}
\tabletypesize{\scriptsize}
\tablecolumns{13}
\tablewidth{0pt}
\tablecaption{ Source Parameters from \textit{wavdetect}}
\tablehead{ 
\colhead{}      & \multicolumn{4}{c}{Soft band (0.3-2.0 keV)} &  
\multicolumn{4}{c}{Medium band (2.0-4.0 keV)}  &
\multicolumn{4}{c}{Hard band (4.0-10.0 keV)} \\ 
\cline{2-5} \cline{6-9} \cline{10-13} 
\colhead{Src }	&\colhead{net counts }
&\colhead{Bkg} &\colhead{S/N} &\colhead{ext } &\colhead{net counts }
&\colhead{Bkg} &\colhead{S/N} &\colhead{ext } &\colhead{net counts }
&\colhead{Bkg} &\colhead{S/N} &\colhead{ext }\\
\colhead{}	&\colhead{$\pm$errors }
&\colhead{} &\colhead{} &\colhead{($\times$PSF) } 	&\colhead{$\pm$errors }
&\colhead{} &\colhead{} &\colhead{($\times$PSF) }	&\colhead{$\pm$errors }
&\colhead{} &\colhead{} &\colhead{($\times$PSF) }\\
\colhead{ (1) }	&\colhead{ (2)}
&\colhead{(3)} &\colhead{(4)} &\colhead{(5)} &\colhead{(6)}
&\colhead{(7)} &\colhead{(8)} &\colhead{(9)} &\colhead{(10)}
&\colhead{(11)} &\colhead{(12)} &\colhead{(13)}}
\startdata
\typeout{Yo}
1	      & $	106.1	\pm	10.5	$ &	3.9	&	33.6		&	0.9	& $	12.8	\pm	3.6	$ &	0.2	&	6.4	&	0.7	& $	6.2	\pm	2.6	$ &	0.8	&	2.7	&	0.5	\\ 
2	      & $	111.1	\pm	11.0	$ &	10.9	&	25.2		&	0.8	& $	17.0	\pm	4.2	$ &	1.0	&	7.3	&	1.0	& $				$ &		&		&		\\ 
3	      & $	206.4	\pm	14.8	$ &	13.6	&	43.2		&	0.8	& $	18.9	\pm	4.5	$ &	1.1	&	8.1	&	0.9	& $				$ &		&		&		\\ 
4	      & $	36.1	\pm	6.5	$ &	5.9	&	10.1		&	1.1	& $	4.8	\pm	2.2	$ &	0.2	&	2.4	&	0.5	& $				$ &		&		&		\\ 
5	      & $	63.0	\pm	8.6	$ &	11.0	&	14.2		&	1.5	& $				$ &		&		&		& $				$ &		&		&		\\ 
6	      & $	32.5	\pm	6.2	$ &	6.5	&	8.8		&	1.2	& $				$ &		&		&		& $				$ &		&		&		\\ 
7	      & $	31.0	\pm	6.2	$ &	8.0	&	7.8		&	1.1	& $				$ &		&		&		& $				$ &		&		&		\\ 
8	      & $	10.1	\pm	3.6	$ &	2.9	&	3.5		&	0.6	& $				$ &		&		&		& $				$ &		&		&		\\
\enddata
\end{deluxetable}


\makeatletter
\def\jnl@aj{AJ}
\ifx\revtex@jnl\jnl@aj\let\tablebreak=\nl\fi
\makeatother
\ptlandscape
\begin{deluxetable}{lccccccccccc}
\tabletypesize{\scriptsize}
\rotate
\tablecolumns{12}
\tablewidth{0pt}
\tablecaption{ Spatial Properties of Bright Sources}
\tablehead{\colhead{ Source }	& \multicolumn{2}{c}{Fit with model PSF}	&\multicolumn{4}{c}{Fit with single Gaussian}	&
\multicolumn{4}{c}{ Fit with Gaussian + constant} & \colhead{Deprojected$^1$ } \\
\colhead{  }	& \colhead{ PSF }& \colhead{ PSF + const}	&\multicolumn{2}{c}{Source }	&
\multicolumn{2}{c}{PSF } & \multicolumn{2}{c}{Source}	&
\multicolumn{2}{c}{PSF } & \colhead{Source size}\\
\colhead{ }	&\colhead{\x2 (dof)}  &\colhead{\x2 (dof)} &\colhead{FWHM}  &\colhead{\x2 (dof)}
& \colhead{FWHM}  &\colhead{\x2 (dof)} &\colhead{FWHM}  &\colhead{\x2 (dof)} &\colhead{FWHM}  &\colhead{\x2 (dof)} \\
\colhead{ }	&\colhead{}  &\colhead{} &\colhead{pixels}  &\colhead{}
& \colhead{pixels}  &\colhead{} &\colhead{pixels}  &\colhead{}
&\colhead{pixels}  &\colhead{} &\colhead{arcsec (pc)}\\
\colhead{ (1) }	&\colhead{ (2)} &\colhead{(3)} &\colhead{(4)}
&\colhead{(5)} &\colhead{ (6)} &\colhead{(7)} &\colhead{(8)} &\colhead{(9)}&\colhead{(10)} &\colhead{(11)}&\colhead{(12)}}
\startdata
1  & 7.60 (4) & 7.51 (3) &$ 3.20_{-0.19}^{+0.20}$&  2.9  (4) &$ 3.65_{-0.23}^{+0.24} $& 1.6 (4) &$ 3.16_{-0.19}^{+0.20}$ &  3.2  (3)&$ 3.58_{-0.24}^{+0.24}$& 1.0 (3)&	p\\[0.1cm]
2 & 2.49 (5) & 2.51 (4) &$ 5.72_{-0.37}^{+0.38}$ &  1.3  (4) &$ 6.09_{-0.37}^{+0.38}$&  1.5  (4) &$5.49_{-0.45}^{+0.45}$&   1.3  (4) &$5.90_{-0.55}^{+0.55}$& 1.3 (3)&	p\\[0.1cm]
3 & 10.4 (8) &  10.4 (7) & $ 5.57 _{-0.29}^{+ 0.30}$ & 5.92 (7) &  $ 6.10 _{-0.32}^{+ 0.33}$ & 4.47 (7)  \\[0.1cm]
5 & 211.2 (5)& 6.7 (4)  &$15.55_{-1.84}^{+2.80}$ &  1.9  (4) &$ 3.36_{-0.25}^{+0.26}$&  0.7  (4) &$7.29_{-3.00}^{+3.00}$&   0.9  (3) &$3.34_{-0.26}^{+0.26}$& 0.5 (3)& $3.17^{+1.54}_{-1.7  }$ ($443 ^{+210 }_{-238}$)\\[0.1cm]
6 & 191.3 (7)& 27.7 (6) &$15.84_{-1.19}^{+1.48}$ &  5.8  (6) &$ 3.28_{-0.36}^{+0.37}$&  0.2  (6) &$10.2_{-1.56}^{+2.26}$&   2.3  (5) &$3.26_{-0.37}^{+0.37}$& 0.1 (5)& $4.73^{+1.1 }_{-0.73 }$ ($662 ^{+154 }_{-102}$)\\[0.1cm]
7 & 159.7 (5)& 2.7 (4)  &$31.28_{-10.9}^{+10.9}$ & 2.2  (4) &$ 3.45_{-0.43}^{+0.47}$&  0.13 (4) &$31.1_{-13.0}^{+13.0}$&  2.2  (3) &$3.40_{-0.45}^{+0.45}$& 0.1 (3)& $15.1^{+6.4 }_{-6.4  }$ ($2114^{+896 }_{-336}$)\\[0.1cm]
10& 123.9 (7)& 8.2 (6)  &$24.81_{-4.33}^{+9.05}$ &  7.7  (6) &$ 2.99_{-0.60}^{+0.78}$&  0.14 (6) &$2.96_{-0.87}^{+36.7}$&  8.4  (5) &$2.97_{-0.63}^{+0.63}$& 0.14 (5)& $12.0^{+4.6 }_{-2.1  }$ ($1687^{+644 }_{-294}$)\\[0.1cm]
11 & 20.6 (7) & 19.8 (6) & $ 2.81 _{-0.08}^{+ 0.08}$ & 22.13 (4)  & $ 3.10 _{-0.09}^{+ 0.10}$ & 21.99 (6)  & $ 2.74 _{-0.07}^{+ 0.08}$ & 4.92 (5)  & $ 3.03 _{-0.09}^{+ 0.09}$ & 14.01 (5) &	p\\
\enddata
\tablenotetext{1}{ Point-like sources are indicated by ``p''.}
\end{deluxetable}


\makeatletter
\def\jnl@aj{AJ}
\ifx\revtex@jnl\jnl@aj\let\tablebreak=\nl\fi
\makeatother
\ptlandscape
\begin{deluxetable}{lccccccccc}
\tabletypesize{\scriptsize}
\tablecaption{ Long Term Variability (ROSAT - {\textit{Chandra}})}
\tablecolumns{6}
\tablewidth{0pt}
\tablehead{ 
\colhead{HRI source$^{1}$} & \colhead{Chandra source(s)}   & \colhead{source
radius}   & \colhead{HRI luminosity$^{2}$}&  
\colhead{Chandra luminosity$^{2}$}& \colhead{Significance } \\
\colhead{} & \colhead{}   & \colhead{(arcsec)}   & \colhead{($10^{39}$ erg/s)} &  
\colhead{($10^{39}$ erg/s)} & \colhead{($\sigma$)} \\
\colhead{ (1) }	&\colhead{ (2)}
&\colhead{(3)} &\colhead{(4)} &\colhead{(5)}&\colhead{(6)} }
\startdata
X-3	&	16	&	5.9	& $	2.89	\pm	0.80	$ & $	7.24	\pm	0.20	$ &5.4\\
X-4	&	22,23,26,30,18.25	&	11.1	& $	8.39	\pm	1.27	$ & $	6.29	\pm	0.19	$ &1.6\\
X-8	&	32, (30,31)$^{3}$	&	6.6	& $	2.77	\pm	0.80	$ & $	3.15	\pm	0.13	$ &0.5\\
X-10	&	41, (43)$^{3}$	&	5.9	& $	2.04	\pm	0.71	$ & $	0.76	\pm	0.07	$ &1.8\\
X-11	&	42	&	5.9	& $	3.54	\pm	0.85	$ & $	6.73	\pm	0.19	$ &3.7\\
X-12	&	44,46	&	5.9	& $	2.66	\pm	0.78	$ & $	5.84	\pm	0.18	$ &4.1\\
\enddata
\tablenotetext{1}{From Fabbiano et al, 1997.}
\tablenotetext{2}{In the 0.1-2.5 keV band. }
\tablenotetext{3}{These sources are at the edge of the extraction region.}
\end{deluxetable}


\makeatletter
\def\jnl@aj{AJ}
\ifx\revtex@jnl\jnl@aj\let\tablebreak=\nl\fi
\makeatother
\ptlandscape
\begin{deluxetable}{lccccccccccccc}
\tabletypesize{\scriptsize}
\rotate
\tablecolumns{11}
\tablewidth{0pt}
\tablecaption{ Single Component Spectral Fits}
\tablehead{ 
\colhead{}     & \multicolumn{2}{c}{Power-law (fixed $\rm{N_{H}}$)} &
\multicolumn{3}{c}{Power-law (free $\rm{N_{H}}$)} &
\multicolumn{2}{c}{RS (fixed $\rm{N_{H}}$)} &   \multicolumn{3}{c}{RS (free $\rm{N_{H}}$)} \\ 
\colhead{Src}	&\colhead{$\Gamma$}
&\colhead{ \x2 (dof)}	& \colhead{$\Gamma$}&
\colhead{$\rm{N_{H}^{1}}$}  &\colhead{\x2 (dof)} & \colhead{kT
(keV)} & \colhead{ \x2 (dof)} &\colhead{kT
(keV)} &\colhead{$\rm{N_{H}^{1}}$}   & \colhead{ \x2 (dof)} \\
\colhead{}	&\colhead{}
&\colhead{}	& \colhead{}&
\colhead{$10^{22}~\rm{cm^{-2}}$}  &\colhead{} & \colhead{keV} 
 & \colhead{} &\colhead{keV} &\colhead{$10^{22}~\rm{cm^{-2}}$}   & \colhead{} \\
\colhead{ (1) }	&\colhead{ (2)}
&\colhead{(3)} &\colhead{(4)} &\colhead{(5)} &\colhead{(6)}
&\colhead{(7)} &\colhead{(8)} &\colhead{(9)}&\colhead{(10)} &\colhead{(11)} }
\startdata
1	& $	1.09	_{-	0.17	}^{+	0.17	} $ &	8.5	(9)	& $	1.61	_{-	0.41	}^{+	0.51	} $ & $	0.13	_{-	0.09	}^{+	0.13	} $ &	2.8	(8)	& $	33.9	(		>6.96		) $ &	5.2	(9)   & $	7.47	_{-	3.55	}^{+	56.53	} $ & $	0.09	_{-	0.06	}^{+	0.07	} $ &	2.8	(8)	\\[0.1cm]
2	& $	1.19	_{-	0.19	}^{+	0.19	} $ &	14.1	(9)	& $	1.79	_{-	0.45	}^{+	0.61	} $ & $	0.14	_{-	0.10	}^{+	0.13	} $ &	7.7	(8)	& $	11.7	_{-	7.33	}^{+	52.3	} $ &	9.9	(9)   & $	6.79	_{-	3.18	}^{+	42.21	} $ & $	0.08	_{-	0.05	}^{+	0.07	} $ &	8.4	(8)	\\[0.1cm]
3	& $	1.46	_{-	0.14	}^{+	0.13	} $ &	31.5	(14)	& $	2.28	_{-	0.37	}^{+	0.48	} $ & $	0.16	_{-	0.07	}^{+	0.09	} $ &	11.8	(13)	& $	3.52	_{-	1.04	}^{+	1.63	} $ &	12.1	(14)  & $	3.45	_{-	0.99	}^{+	1.59	} $ & $	0.04	_{-	0.03	}^{+	0.04	} $ &	11.9	(13)	\\[0.1cm]
5	& $	2.42	_{-	0.16	}^{+	0.17	} $ &	97.2	(18)	& \bf$	7.22	_{-	1.71	}^{+	2.39	} $ & \bf$	0.73	_{-	0.26	}^{+	0.36	} $ &\bf	42.4	(17)	& $	0.76	_{-	0.11	}^{+	0.05	} $ &	67.8	(18)  & $	0.77	_{-	0.06	}^{+	0.05	} $ & $	0.00	_{-	0.00	}^{+	0.01	} $ &	61.3	(17)	\\[0.1cm]
6	& $	2.22	_{-	0.26	}^{+	0.29	} $ &	40.2	(13)	& $	4.95	_{-	1.51	}^{+	3.64	} $ & $	0.37	_{-	0.21	}^{+	0.54	} $ &	25.0	(12)	& $	0.32	_{-	0.04	}^{+	0.05	} $ &	10.5	(13)  & $	0.33	_{-	0.05	}^{+	0.05	} $ & $	0.00	_{-	0.00	}^{+	0.16	} $ &\bf	10.1	(12)	\\[0.1cm]
10	& $	2.11	_{-	0.27	}^{+	0.30	} $ &	37.8	(14)	& $	10	_{-	4.06	}^{+	0.00	} $ & $	1.17	_{-	0.09	}^{+	0.09	} $ &	16.6	(13)	& $	0.83	_{-	0.08	}^{+	0.06	} $ &	15.5	(14)  & $	0.84	_{-	0.07	}^{+	0.05	} $ & $	0.00	_{-	0.00	}^{+	0.04	} $ &\bf	12.9	(13)	\\[0.1cm]
11	& $	0.84	_{-	0.07	}^{+	0.07	} $ &	103.3	(45)	& $	1.54	_{-	0.18	}^{+	0.19	} $ & $	0.19	_{-	0.05	}^{+	0.06	} $ &	36.5	(44)	& $				>47.0		 $ &	105.3	(45)  & $	11.40	_{-	4.56	}^{+	25.00	} $ & $	0.16	_{-	0.04	}^{+	0.04	} $ &	38.6	(44)	\\
\enddata
\end{deluxetable}


\makeatletter
\def\jnl@aj{AJ}
\ifx\revtex@jnl\jnl@aj\let\tablebreak=\nl\fi
\makeatother
\ptlandscape
\begin{deluxetable}{lccccccccccccc}
\tabletypesize{\scriptsize}
\rotate
\tablecolumns{11}
\tablewidth{0pt}
\tablecaption{Luminosities of Sources with Spectral Fits}
\tablehead{ 
\colhead{Source} & \colhead{Rate}      & \multicolumn{2}{c}{Flux (observed)$^{1}$} &  
\multicolumn{2}{c}{Flux (corr. Gal.)$^{1}$} & \multicolumn{2}{c}{Flux (corr. tot.)$^{1}$} &\multicolumn{2}{c}{Lumin (observed)$^{2}$} &  
\multicolumn{2}{c}{Lumin (corr. Gal.)$^{2}$} & \multicolumn{2}{c}{Lumin (corr. tot.)$^{2}$}\\ 
\colhead{}	&\colhead{$10^{-2}\rm{cts/s}$}
&\colhead{soft} &\colhead{hard} &\colhead{soft}
&\colhead{hard}&\colhead{soft} &\colhead{hard}   &\colhead{soft}
&\colhead{hard} &\colhead{soft} &\colhead{hard}&\colhead{soft} &\colhead{hard}\\
\colhead{ (1) }	&\colhead{ (2)}
&\colhead{(3)} &\colhead{(4)} &\colhead{(5)} &\colhead{(6)}
&\colhead{(7)} &\colhead{(8)} &\colhead{(9)} & \colhead{(10)}&\colhead{(11)} &\colhead{(12)} &\colhead{(13)} & \colhead{(14)}}
\startdata
\scriptsize
\typeout{Yo}
1  &  0.0020  &0.57  &1.09  &  0.62  &  1.09  &  1.08  &  1.09  &38.76  &  39.04  &  38.79  &  39.04  &  39.04  &  39.04  \\
2  &  0.0019  &0.53  &0.71  &  0.60  &  0.72  &  1.09  &  0.72  &38.72  &  38.86  &  38.78  &  38.86  &  39.04  &  38.86  \\
3  &  0.0032  &0.90  &0.86  &  1.02  &  0.87  &  2.17  &  0.87  &38.96  &  38.94  &  39.01  &  38.94  &  39.34  &  38.94  \\
5  &  0.0039  &1.04  &0.03  &  1.38  &  0.03  &  7.49  &  0.03  &39.02  &  37.46  &  39.14  &  37.46  &  39.88  &  37.47  \\
6  &  0.0022  &0.60  &0.09  &  0.81  &  0.09  &  1.23  &  0.09  &38.78  &  37.95  &  38.91  &  37.95  &  39.09  &  37.95  \\
10  &  0.0022  &0.56  &0.01  &  0.76  &  0.01  &  745.50  &  0.01  &38.75  &  37.06  &  38.88  &  37.07  &  41.88  &  37.07  \\
11  &  0.0103  &2.96  &6.37  &  3.31  &  6.38  &  5.92  &  6.45  &39.47  &  39.81  &  39.52  &  39.81  &  39.77  &  39.81  \\
12  &  0.0009  &0.28  &0.65  &  0.29  &  0.65  &  1.21  &  0.67  &38.45  &  38.82  &  38.46  &  38.82  &  39.09  &  38.83  \\
\enddata
\tablenotetext{1}{ Flux in units of $10^{-14}$\ergs. Soft:  0.1-2.5keV band, hard: 2.5-10.0keV band.}
\tablenotetext{2}{Log of the luminosity in units of \ergs. }
\end{deluxetable}


\makeatletter
\def\jnl@aj{AJ}
\ifx\revtex@jnl\jnl@aj\let\tablebreak=\nl\fi
\makeatother
\ptlandscape
\begin{deluxetable}{lcccccccccccccc}
\tabletypesize{\scriptsize}
\rotate
\tablecolumns{13}
\tablewidth{0pt}
\tablecaption{ Double Component Spectral Fits }
\tablehead{\colhead{ }	&\multicolumn{4}{c}{ Simultaneous PO+RS fits}
&\multicolumn{4}{c}{PO fits above 2.0~keV} &
\multicolumn{4}{c}{Additional \nh~component in PO fit} \\ 
\cline{2-5}  \cline{6-10}
\colhead{Source}  &\colhead{$\Gamma$ }  &\colhead{
kT (keV)} & \colhead{ $\rm{N_{H}^{1}}$}
&\colhead{\x2 (dof) } &\colhead{$\Gamma$ }  & \colhead{kT (keV)} & \colhead{ $\rm{N_{H}^{2}}$}  & \colhead{\x2/(dof) }
& \colhead{kT (keV)} & \colhead{ $\rm{N_{H}^{hard,1}}$}& \colhead{ $\rm{N_{H}^{tot,2}}$}
&\colhead{\x2/(dof) }\\
\cline{2-6}  \cline{7-12}
\colhead{ (1) }	&\colhead{ (2)}
&\colhead{(3)} &\colhead{(4)} &\colhead{(5)} & \colhead{ (6) }	&\colhead{ (7)}
&\colhead{(8)} &\colhead{(9)} &\colhead{(10)} &\colhead{(11)}&\colhead{(12)} &\colhead{(13)}}
\startdata
2	&$	1.75	_{-0.37	}^{+0.28	}$ & $ 0.82	(>0.1)			$ & $ 0.12	_{-0.08	}^{+0.13	}$ &	6.4	(6)	&  &  &  &   &    &    \\[0.1cm]                                                                                                                                                                                                                                                                                                        
3	&$	2.01	_{-0.54	}^{+0.54	}$ & $ 1.30	_{-0.27	}^{+3.65	}$ & $ 0.09	_{-0.08	}^{+0.11	}$ &	7.7	(11)	& $ 2.14_{-0.67}^{+0.82  }$     & $ 1.26_{-0.26}^{+2.06}$     & $0.12 _{-0.04 }^{+0.04 }$ & 8.0/12\\    [0.1cm]
5	&$	3.75	_{-0.34	}^{+1.11	}$ & $ 0.66	_{-0.10	}^{+0.11	}$ & $ 0.10	_{-0.10	}^{+0.03	}$ &\bf	20.7	(15)	                                                                   \\                                                                           [0.1cm]                                                                                                                                                                                 
6	&$	2.48	_{-1.36	}^{+1.26	}$ & $ 0.34	_{-0.06	}^{+0.09	}$ & $ 0.04	_{-0.04	}^{+1.14	}$ &\bf	2.4	(10)	                                                                   \\                                                                           [0.1cm]                                                                                                                                                                                 
10	&$	>5.17				$  & $ 0.83	_{-0.12	}^{+0.07	}$ & $ 0.28	_{-0.17	}^{+0.19	}$ &	4.0	(11)	                                                                   \\
\enddata
\tablenotetext{1}{$\rm{N_{H}}$ in units of of $10^{22}~\rm{cm^{-2}}$.}
\tablenotetext{1}{Column density for PO component. }
\tablenotetext{2}{Column density for overall (PO+RS) model. }
\tablenotetext{3}{Unconstrained parameter. }
\end{deluxetable}


\makeatletter
\def\jnl@aj{AJ}
\ifx\revtex@jnl\jnl@aj\let\tablebreak=\nl\fi
\makeatother
\ptlandscape
\begin{deluxetable}{lccccccccc}
\tabletypesize{\scriptsize}
\tablecolumns{5}
\tablewidth{0pt}
\tablecaption{Fits of Emission Lines}
\tablehead{ 
\colhead{Source} & \colhead{Energy}   & \colhead{Line ID}   &
\colhead{Norm} & \colhead{\x2 (dof)}  \\ 
\colhead{}	&\colhead{(keV)} & \colhead{}& \colhead{$\rm{10^{7}phot/cm^{2}/s}$} \\
\colhead{ (1) }	&\colhead{ (2)}
&\colhead{(3)} &\colhead{(4)} &\colhead{(5)} }
\startdata
11	& $	1.50_{-0.04}^{+0.02}$ &	Mg ${\sc XIII}$	& $	4.10_{-2.88}^{+3.26	}$& 	31.1 (41)	\\[0.1cm]
\hline																
16	& $	0.45_{-0.02}^{+0.04}$ &	N ${\sc VI}$	& $	23.00_{-19.74}^{+26.20	}$& 	103.4 (101)	\\[0.1cm]
	& $	1.33_{-0.02}^{+0.02}$ &	Mg ${\sc XI}$	& $	6.80_{-4.18}^{+5.40	}$& 		\\[0.1cm]
	& $	2.17_{-0.06}^{+0.06}$ &	?	& $	9.60_{-6.00}^{+6.10}	$& 		\\[0.1cm]
\hline																	
42	& $	1.23_{-0.03}^{+0.02}$ &	Na ${\sc XI}$	& $	6.43_{-3.73}^{+3.67	}$& 	47.3 (74)	\\[0.1cm]
	& $	1.02_{-0.03}^{+0.02}$ &	Ne ${\sc X}$	& $	7.00_{-3.70}^{+5.20	}$& 		\\[0.1cm]
\enddata
\end{deluxetable}

%% file: ms.bbl
\begin{thebibliography}{}

\bibitem[Arnaud 1996]{}Arnaud K., 1996, in Astronomical Data Analysis
Software and Systems V,  ASP Conf. Series volume 101, eds. G. Jacoby
 \& J. Barnes 




\bibitem[Dobr ]{} Dobrzycki, A., Ebeling, H., Glotfelty, K., Freeman, P., Damiani, F., Elvis, M., Calderwood, T., 2000, \chandra Detect 1.0 User Guide, http://asc.harvard.edu/ciao/documents\_manuals.html


\bibitem[Fabian 1996]{} Fabian A., \& Terlevich R., 1996  MNRAS,  280, 5

\bibitem[Fabbiano 1989]{f89} Fabbiano, G. 1989, Ann. Rev. Ast. Ap., 27, 87

\bibitem[Fabbiano 1995]{f95} Fabbiano, G. 1995, in X-ray Binaries, ed.\ W. H. G. Lewin, J. van Paradijs, \& E. P. J. van den Heuvel (Cambridge: University Press), p.\ 390

\bibitem[Fabbiano \etal\ 1982]{fabb82} Fabbiano, G., Feigelson, E., \&
Zamorani, G. 1982, \apj, 256, 397

\bibitem[Fabbiano \& Trinchieri 1983]{fabb83} Fabbiano, G. \& Trinchieri,
G.  1983, \apj, 266, L5

\bibitem[Fabbiano \etal\ 1997]{}Fabbiano, G., Schweizer, F., \& Mackie, G.
1997, \apj, 478, 542

\bibitem[Fabbiano \etal\ 2001]{} Fabbiano, G., Zezas, A., \& Murray, S.
2001, \apj, 554, 1035 (Paper~I)




\bibitem[Freeman etal 2001]{} Freeman P.E., Kashyap V., Rosner R. \&
Lamb D.Q., 2001a, astro-ph/0108429

\bibitem[Freeman etal 2001]{} Freeman P.E., Doe S. \& Siemiginowska
A., 2001b, astro-ph/0108426


 K.~Y., Lee, S.-W., \& Lee, T.-H.\ 2001, \apj, 548, 172 

\bibitem[Garmire 1997]{}Garmire, G. P. 1997, AAS, 190, 3404

\bibitem[Gehrels(1986)]{1986ApJ...303..336G} Gehrels, N.\ 1986, \apj, 303, 
336 


\bibitem[Giacconi et al.(2001)]{2001ApJ...551..624G} Giacconi, R.~et al.\ 
2001, \apj, 551, 624 


















\bibitem[Nagase(1989)]{1989PASJ...41....1N} Nagase, F.\ 1989, \pasj, 41, 1 

\bibitem[Neff \& Ulvestad(2000)]{2000AJ....120..670N} Neff, S.~G.~\& 
Ulvestad, J.~S.\ 2000, \aj, 120, 670 


\bibitem[Plewa 1995]{pl95} Plewa T.,   1995, \mnras, 275, 143


\bibitem[Read \etal\ 1995]{read95} Read, A. M., Ponman, T. J., \&
Wolstencroft, R. D.  1995, \mnras, 277, 397


\bibitem[Sansom \etal\ 1996]{sans96} Sansom, A.E., Dotani, T., Okada, K.,
Yamashita, A., \& Fabbiano, G.  1996, \mnras, 281, 48

\bibitem[Schlegel 1995]{}Schlegel E., 1995, Reports of Progress in
Physics, 58, 1375

\bibitem[Stark et al.(1992)]{1992ApJS...79...77S} Stark, A.~A., Gammie, C.~F., Wilson, R.~W., Bally, J., Linke, R.~A., Heiles, C., \& Hurwitz, M.\ 1992, \apjs, 79, 77


\bibitem[Tan Lew 1995]{f95} Tanaka, F. \& Lewin W. 1995, in X-ray Binaries,
ed.\ W. H. G. Lewin, J. van Paradijs, \& E. P. J. van den Heuvel
(Cambridge: University Press), p.\ 126


\bibitem[Terl 1994]{t95} Terlevich, R. 1994, in Circumstellar Media in
the Late Stages of Stellar Evolution,
ed.\ R.E.S. Clegg, I.R. Stevens,  W.P.S Meikle, J. van Paradijs, 
(Cambridge: University Press), p.\ 153

\bibitem[Toomre \& Toomre(1972)]{1972ApJ...178..623T} Toomre, A.~\& Toomre, J.\ 1972, \apj, 178, 623


\bibitem[van Paradijs 1999]{1995ApJ...447L..33V} Van Paradijs, 
J., 1999, in the Many Faces of Neutron Stars, ed.\ R. Buccheri, J. van Paradijs, M.A.
     Alpar (Kluwer Academic Publishers), p.\ 279  (astro-ph/9802177)


\bibitem[Van Speybroeck \etal\ 1997]{} Van Speybroeck, L., Jerius D., Edgar, R. J., Gaetz, T. J., Zhao, P. \& Reid, P. B.1997, Proc. SPIE 3113, 89 



\bibitem[Weisskopf \etal\ 2000]{}Weisskopf, M., Tananbaum, H., Van Speybroeck, L. \& O'Dell, S.
2000, Proc. SPIE 4012 (astro-ph 0004127)






\bibitem[Zezas \etal\ 2002]{} Zezas, A. \& Fabbiano, G.,  2001, submitted to \apj (Paper~IV)

\bibitem[Zezas \etal\ 2002]{} Zezas, A., Fabbiano G., Rots, A. H. \&
Murray S. S,  2001, submitted to \apj (Paper~III)

\end{thebibliography}
